\documentclass[twocolumn]{aastex63}
\usepackage{amsmath}

\usepackage{xcolor, fontawesome, microtype}

\usepackage{ mathrsfs }
\usepackage[T1]{fontenc}
\usepackage{xcolor}

\received{\today}
\revised{\today}
\accepted{\today}
\submitjournal{AJ}

\newcommand{\teff}{$T_\textrm{eff}\;$}
\newcommand{\Kepler}{\emph{Kepler}\;}
\newcommand{\Ktwo}{\emph{K2}\;}
\newcommand{\tess}{\emph{TESS}\;}
\newcommand{\GAIA}{\emph{Gaia}\;}

\shorttitle{\Ktwo vs.\ \Kepler}

\shortauthors{Zink et al. (2023)}

\begin{document}

\title{Scaling \emph{K2}. VI. Reduced Small Planet Occurrence in High Galactic Amplitude Stars}

\correspondingauthor{Jon Zink}
\email{jzink@caltech.edu}

\author[0000-0003-1848-2063]{Jon K.\ Zink}
\altaffiliation{NHFP Sagan Fellow}
\affiliation{Department of Astronomy, California Institute of Technology, Pasadena, CA 91125}

\author[0000-0003-3702-0382]{Kevin K. Hardegree-Ullman}
\affiliation{Steward Observatory, The University of Arizona, Tucson, AZ 85721}

\author[0000-0002-8035-4778]{Jessie L. Christiansen}
\affiliation{Caltech/IPAC-NASA Exoplanet Science Institute, Pasadena, CA 91125}

\author[0000-0003-0967-2893]{Erik A. Petigura}
\affiliation{Department of Physics and Astronomy, University of California, Los Angeles, CA 90095}

\author[0000-0001-8153-639X]{Kiersten M. Boley}
\altaffiliation{NSF Graduate Research Fellow}
\affiliation{Department of Astronomy, The Ohio State University, Columbus, OH 43210, USA}

\author[0000-0002-6673-8206]{Sakhee Bhure}
\altaffiliation{Volunteer Researcher}
\affiliation{Caltech/IPAC-NASA Exoplanet Science Institute, Pasadena, CA 91125}

\author[0000-0002-7670-670X]{Malena Rice}
\altaffiliation{51 Pegasi b Fellow}
\affiliation{Department of Physics and Kavli Institute for Astrophysics and Space Research, MIT, Cambridge, MA 02139, USA}
\affiliation{Department of Astronomy, Yale University, New Haven, CT 06511, USA}

\author[0000-0001-7961-3907]{Samuel W.\ Yee}
\affiliation{Department of Astrophysical Sciences, Princeton University, 4 Ivy Lane, Princeton, NJ 08544, USA}

\author[0000-0002-0531-1073]{Howard Isaacson}
\affiliation{501 Campbell Hall, University of California at Berkeley, Berkeley, CA 94720, USA}
\affiliation{Centre for Astrophysics, University of Southern Queensland, Toowoomba, QLD, Australia}

\author[0000-0002-3853-7327]{Rachel B. Fernandes}
\affil{Lunar and Planetary Laboratory, The University of Arizona, Tucson, AZ 85721, USA}
\affil{Alien Earths Team, NASA Nexus for Exoplanet System Science, USA}

\author[0000-0001-8638-0320]{Andrew W.\ Howard}
\affiliation{Department of Astronomy, California Institute of Technology, Pasadena, CA 91125, USA}

\author[0000-0002-3199-2888]{Sarah Blunt}
\affiliation{Department of Astronomy, California Institute of Technology, Pasadena, CA 91125, USA}

\author[0000-0001-8342-7736]{Jack Lubin}
\affiliation{Department of Physics \& Astronomy, University of California Irvine, Irvine, CA 92697, USA}

\author[0000-0003-1125-2564]{Ashley Chontos}
\altaffiliation{Henry Norris Russell Fellow}
\affiliation{Department of Astrophysical Sciences, Princeton University, 4 Ivy Lane, Princeton, NJ 08540, USA}

\author[0000-0001-9771-7953]{Daria Pidhorodetska}
\affiliation{Department of Earth and Planetary Sciences, University of California, Riverside, CA, USA}

\author[0000-0003-2562-9043]{Mason G. MacDougall}
\affiliation{Department of Physics and Astronomy, University of California, Los Angeles, CA 90095}

\begin{abstract}

In this study, we performed a homogeneous analysis of the planets around FGK dwarf stars observed by the \Kepler and \Ktwo missions, providing spectroscopic parameters for 310 \Ktwo targets ---including 239 Scaling \Ktwo hosts--- observed with Keck/HIRES. For orbital periods less than 40 days, we found that the distribution of planets as a function of orbital period, stellar effective temperature, and metallicity was consistent between \Ktwo and \Kepler, reflecting consistent planet formation efficiency across numerous $\sim1$ kpc sight-lines in the local Milky Way. Additionally, we detected a 3$\times$ excess of sub-Saturns relative to warm Jupiters beyond 10 days, suggesting a closer association between sub-Saturn and sub-Neptune formation than between sub-Saturn and Jovian formation. Performing a joint analysis of \Kepler and \Ktwo demographics, we observed diminishing super-Earth, sub-Neptune, and sub-Saturn populations at higher stellar effective temperatures, implying an inverse relationship between formation and disk mass. In contrast, no apparent host-star spectral-type dependence was identified for our population of Jupiters, which indicates gas-giant formation saturates within the FGK mass regimes. We present support for stellar metallicity trends reported by previous \Kepler analyses. Using \GAIA DR3 proper motion and RV measurements, we discovered a galactic location trend: stars that make large vertical excursions from the plane of the Milky Way host fewer super-Earths and sub-Neptunes. While oscillation amplitude is associated with metallicity, metallicity alone cannot explain the observed trend, demonstrating that galactic influences are imprinted on the planet population. Overall, our results provide new insights into the distribution of planets around FGK dwarf stars and the factors that influence their formation and evolution.
\end{abstract}



\section{Introduction}
The \Kepler Space Telescope identified over 4700 transiting exoplanet candidates \citep{NEAcumulative} through continuous photometric monitoring of a single patch of the sky \citep{koc10,bor11}. This field was selected because of the predominance of Sun-like stars, in hopes that these stars would yield a meaningful occurrence rate for Earth analogs \citep{bat10}. The consistency of the mission photometry enabled the development of automated search algorithms \citep{jen10} and accurate quantification of the sample biases \citep{pet13a,chr15,dres15,chr17,cou17b,chr20}. Early work found that the small planet populations ($<4R_\Earth$) follow a power law in both radius and period space \citep{you11,how12,pet13b,dres13,mui15,dres15}, suggesting an abundant population of small short-period planets with no solar system equivalent. The final processing of DR25 \citep{tho18} has been used ubiquitously throughout the field to understand the underlying occurrence of exoplanets in our local region of the galaxy. Numerous first-order results have transpired from \Kepler DR25, including high confidence occurrence estimates for the super-Earths, sub-Neptunes, sub-Saturns, and Jupiters within 100 day periods \citep[e.g.,][]{muld18,pet18Met,hsu19,zin19,he19,har19}. These strong results provide an excellent estimate for the local galactic exoplanet population, but the biases imposed by the original field selection have yet to be fully explored. 

After the spacecraft underwent an operational malfunction, the telescope could no longer remain pointed at the \Kepler field. Thoughtful field selection along the ecliptic plane reduced the ongoing solar pressure that caused persistent spacecraft drift, providing an opportunity to probe unique stellar populations and test the robustness of the \Kepler results. The \Ktwo mission was born out of this spacecraft pointing issue and 19 unique fields were observed, representing a more isotropic sampling of the local galaxy \citep{how14, cle16}. Unfortunately, the reaction-wheel malfunction that concluded the primary \Kepler mission riddled the \Ktwo photometry with correlated noise variations, which could not be addressed with the base \Kepler software. This led to numerous community-based efforts to systematically remove these spacecraft artifacts and identify new planet signals \citep{van16b,lug16,pet18c,lug18}. Eventually, nearly 1000 new planets would be identified in \Ktwo photometry through piecemeal efforts \citep{bar16,ada16,cro16,pop16,dres17,liv18,may18,ye18,kru19,zin19c}. However, the absence of a rigorous homogeneous analysis of the photometry made it difficult to carry out robust demographic analyses.

The Scaling \Ktwo series provided the first homogeneously derived stellar \citep{har20} and planet \citep{zin20a,zin21} samples, and the catalog measurements needed to perform robust occurrence rates using the \Ktwo planet sample. We provide further discussion of these samples and the filters used to focus our current study and ensure purity in Sections \ref{sec:stellarSample} and \ref{sec:planetSample} of this paper. In Scaling \Ktwo III we provided early planet occurrence results from our catalog (using only the Campaign 5 sample), showing our pilot study had consistent occurrence measurements with \Kepler \citep{zin20b}. In Scaling \Ktwo V we validated 60 new planet detections \citep{chr22}, providing further evidence that our novel planet detections were robust against astrophysical false positives. In this work we used the full catalog to provide robust planet occurrence results, as dictated by our forward-modeling technique described in Section \ref{sec:FM}, for the entirety of the transit-based \Ktwo campaigns, and we compared these results with an analogous \Kepler demographic analysis in Section \ref{sec:Results}. We also tested established stellar trends and identified a new location based parameter in Section \ref{sec:trends}. We discuss our findings in Section \ref{sec:Disc} and conclude with a summary of our findings in Section \ref{sec:Conc}.  

\begin{figure*}
\centering \includegraphics[width=\textwidth{}]{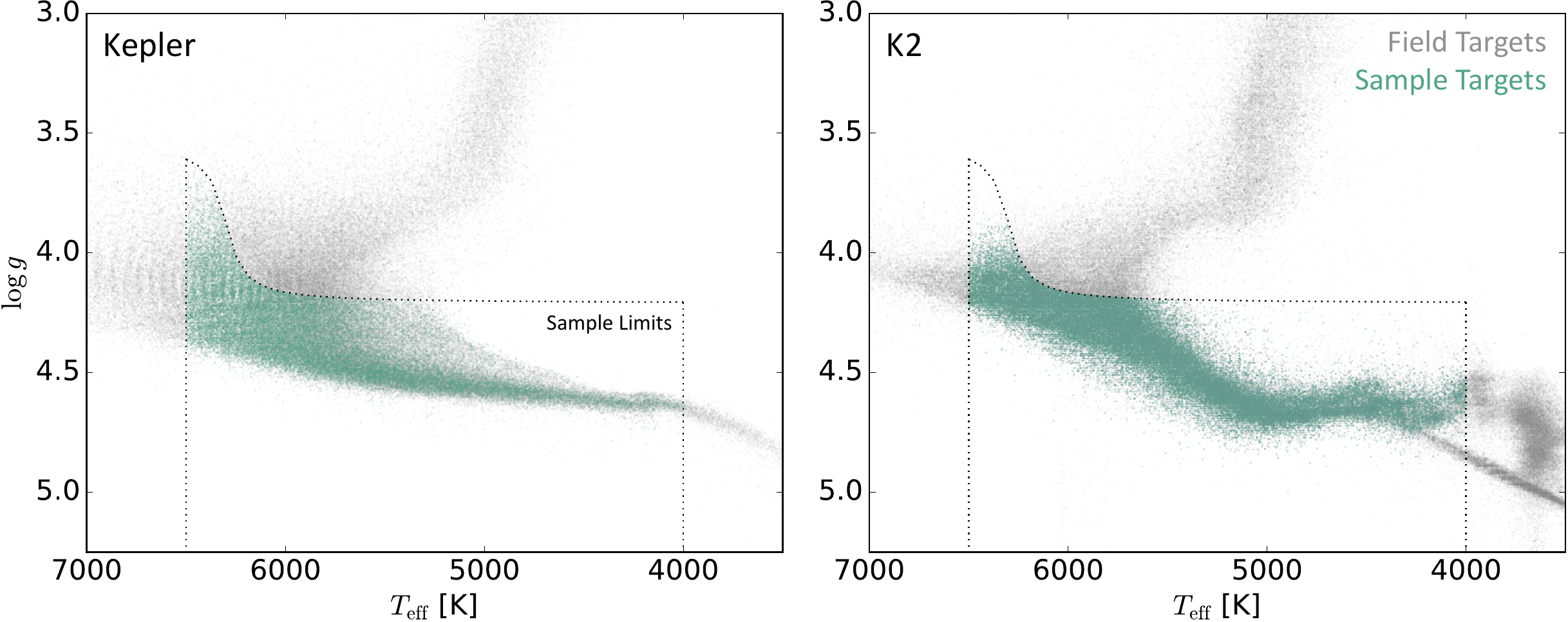}
\caption{The stellar sample from \Kepler and \Ktwo portrayed in the \teff and $\log g$ plane. The gray points show the \Kepler and \Ktwo field sample, while the colored dots highlight the FGK dwarfs selected for this study.
\label{fig:stellar}}
\end{figure*}

\section{The Stellar Sample}
\label{sec:stellarSample}
The focus of this work is to compare planet occurrence between the \Kepler and \Ktwo missions. We chose to limit our study to FGK stars, which were well sampled by the \Ktwo and \Kepler target selection functions. The analysis of incomplete exoplanet samples, as done here, requires a strong grasp on the underlying stellar population the respective planets have been drawn from. For this study we begin with the \cite{ber20} catalog for \Kepler stars and the \cite{har20} catalog for the \Ktwo stellar sample. We acknowledge that these two studies used unique methodologies for deriving stellar parameters; \cite{ber20} used photometric mapping to isochrone models, while \cite{har20} relied on empirically based parameters photometrically optimized through a random forest classifier. The different methodologies may lead to systematic offsets, but \cite{zin20b} tested these two catalogs against the homogeneously derived \GAIA DR2 stellar radius\footnote{Stellar radius measurements have the greatest impact on occurrence measurements because the of sample detection efficiency's strong dependence on the projected stellar area. Systematic effects from other parameterization differences will be minor and are unlikely to produce significant occurrence offsets between the samples.} parameters and found an average offset of $0.06R_{\Sun}$ between these two catalogs, which is comparable to the 5\% uncertainty exhibited by a majority of individual stellar radius measurements.

In an effort to reduce parameter uncertainties, we updated the [Fe/H], $\log g$, and \teff values to reflect the most recent spectroscopic surveys, including LAMOST DR8 \citep{wan22}, APOGEE DR17 \citep{abd22}, the CKS survey \citep{pet17Cat}, 73 existing Keck/HIRES \Ktwo target parameterizations \citep{pet18c} and 237 new \Ktwo target characterizations using Keck/HIRES. \footnote{Existing homogeneous stellar characterization using \GAIA DR3 from \cite{cre22, fou22, and22, ber23} relied on the spacecraft's low-resolution Bp/Rp spectragraph. The constraining power of this instrument appears to be on order 0.2 dex for [Fe/H] in its current form and requires careful offset correction; see Figure 1 of \citet{ber23}. To avoid sample contamination we did not include these parameter updates in our stellar sample.} These new target spectra follow the CKS survey processing protocol and provide the same level of homogeneity and precision; see Appendix \ref{sec:hires} for a detailed account of our new stellar parameters. In testing we found a vast majority of the spectra-based updates are within the previously determined measurement uncertainties, minimizing any systematic changes. In particular, both LAMOST and APOGEE [Fe/H] values exhibited an average 0.02 dex offset when compared to overlapping targets in the CKS catalog, well within the reported 0.03--0.1 dex uncertainty of either of these surveys. Further discussion of these precise parameter values can be found in Section \ref{sec:trends}.

\begin{figure*}
\centering \includegraphics[width=\textwidth{}]{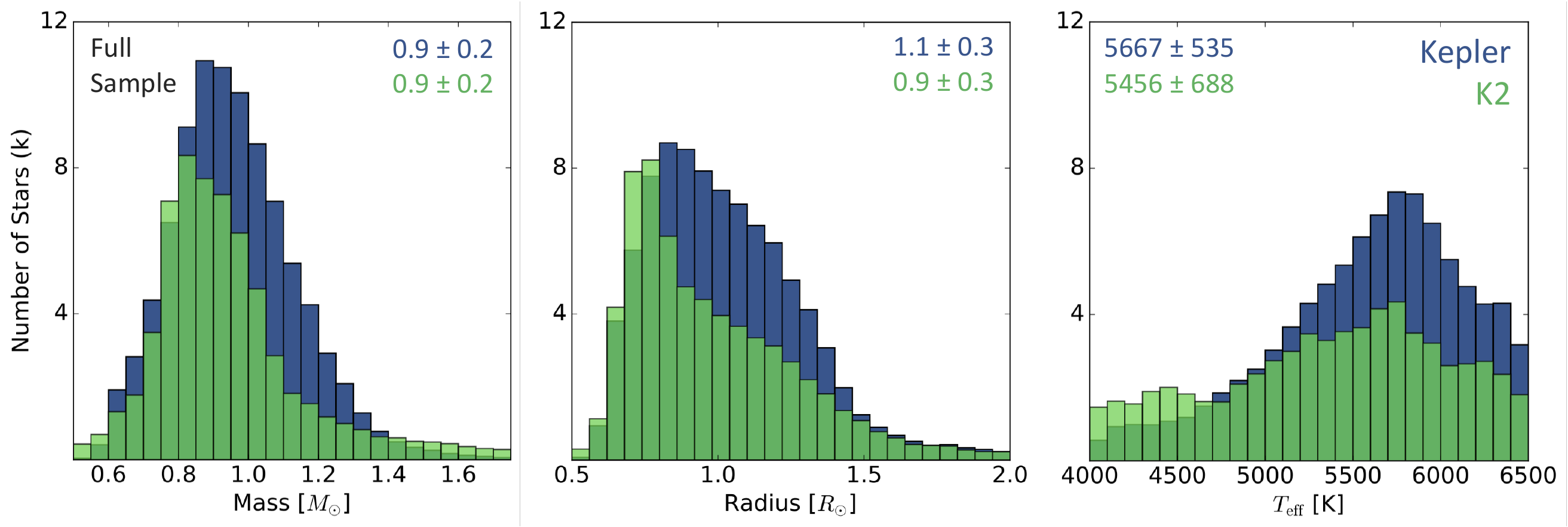}
\caption{A comparative display of the \Kepler and \Ktwo FGK stellar parameters. The y-axis shows the number of stars in each bin in units of 1,000 stars. The median and median absolute deviation values for each distribution are reported in the upper corner of each plot. We provide stellar mass for a population level visual comparison, noting the dominance of loose photometric constraint within our sample.
\label{fig:stellarcomp}}
\end{figure*}

We began with the entirety of both catalogs, comprised of 186,301 \Kepler targets and 222,088 unique \Ktwo targets. We isolated the FGK stars by selecting targets with \teff between 4000 and 6500K, removing 19,697 \Kepler and 52,727 \Ktwo targets. To ensure these targets were dwarfs, we follow the advisement of \citet{hub16} and only selected stars with:
\begin{equation}
\log g\ge\frac{\arctan\bigg(\frac{6300K-\textrm{T}_\textrm{eff}}{67.172K}\bigg)}{4.671}+3.876 \textrm{ dex},
\end{equation}
excluding 61,056 \Kepler stars and 73,963 \Ktwo stars. The 4\arcsec\ pixel width of the spacecraft instrument allows for unresolved stellar companions. We relied on the \GAIA Renormalized Unit Weight Error (RUWE) metric to minimize this potential source of contamination and only included targets with RUWE $<1.4$ (as suggested by \citealt{lin18}). Additionally, \GAIA DR3 provides a flag, "non\_single\_star", which denotes sources that provide evidence of a binary. Removing targets with \GAIA binary flags and high RUWE values, we excluded 14,410 \Kepler and 31,017 \Ktwo targets. The remaining sample may consist of photometrically turbulent targets, which will not yield meaningful candidates and only slow down the occurrence processing. Therefore, we placed an upper bound on the photometric noise metric known as the Combined Differential Photometric Precision (CDPP; \citealt{chr12}). For \Kepler we required $\textrm{CDPP}_\textrm{7.5hr}<1000$ppm, excluding 288 targets, and $\textrm{CDPP}_\textrm{8hr}<1200$ppm for \Ktwo, excluding an additional 11,425 targets. We found 90,850 \Kepler and 64,454 \Ktwo stars met the discussed criteria, and labeled this selection of stars as our Full Sample (in Section \ref{sec:trends} we limit this sample further). The selected catalog of stars is displayed in Figure \ref{fig:stellar} against the background of the total field sample.

\subsection{Kepler and K2 Sample Comparison}
The cuts selected for FGK dwarfs; however, the exact distribution of FGK dwarfs is slightly different between the two missions. In Figure \ref{fig:stellarcomp} we show the relevant stellar parameter distributions for both \Kepler and \Ktwo. Overall, there is a general consistency between both sets of stars. Under more careful inspection, it appears the \Ktwo sample is on average $\sim200 K$ cooler and $\sim0.2R_\Earth$ smaller than the \Kepler stellar sample. We provide a detailed discussion of the dependence of spectral class on planet occurrence in Section \ref{sec:trends}.

\begin{figure*}
\centering \includegraphics[width=\textwidth{}]{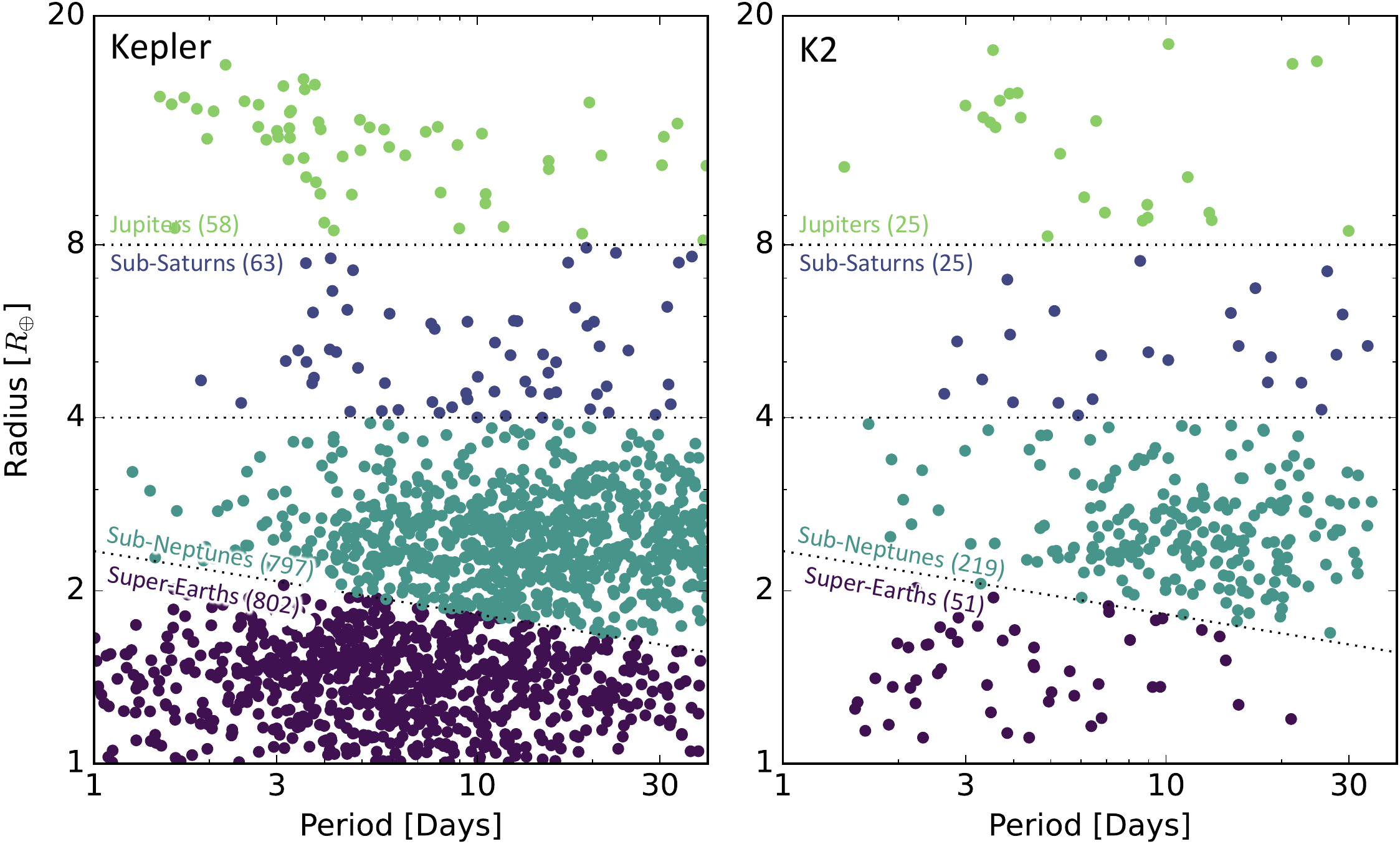}
\caption{The sample of \Kepler and \Ktwo planets selected for this demographic analysis. The gray dotted line separates the planet classes accordingly. The number of planets in each bin is denoted next to the respective class title. We employed the two two-dimensional support-vector machine model from \citet{ho23} for our super-Earth sub-Neptune bound.
\label{fig:planetcomp}}
\end{figure*}

\section{The Planet Sample}
\label{sec:planetSample}
We drew our sample of planets from the two available homogeneous catalogs, \Kepler DR25 \citep{tho18} with updates from \citet{ber20b} and Scaling \Ktwo \citep{zin21}. Both catalogs provide the necessary completeness measurements needed to carry out a robust statistical analysis. Before slicing this sample into the required bins it is worth considering the potential for systematic offsets between the catalog parameters. The three-transit minimum requirement of each catalog reduces period uncertainties to within 0.01\%, rendering any existing offsets negligible. However, the planet radii are subject to significant modification from the necessary data processing schemes, requiring a more detailed examination. \citet{chr15} considered how the \Kepler PDC pipeline impacted the radius parameters by looking for offsets between the injected (known values) and recovered (measured values) planet signals. Overall, the software systematically reduced the planet radius by $1.7\%$, well within the 5\% uncertainty presented in each catalog. The \Ktwo light curves required even further processing, given the consistent drift of the telescope. \citet{zin20a} considered how this additional detrending would impact the measured radius values through an analogous comparison of the injected and recovered signals, finding a $2.3\%$ average reduction in radius parameters. Again, this offset is well within the expected uncertainty. Furthermore, these two radius reduction estimates are in the same direction. Thus, the systematic radius offset between these catalogs is only $\sim0.6\%$, which is negligible and will not impact our overall analysis and comparisons.    

We started with the complete \Kepler DR25 planet candidate sample (4,612 candidates). We then removed 1,647 candidates not hosted by stars in our stellar sample. It is important that our samples span comparable period ranges, so we excluded candidates with periods less than one day (101 candidates) and periods beyond 40 days (667 candidates). To ensure purity and minimize the contamination from eclipsing binaries, we required an impact parameter of $b<0.9$, removing 236 planets from our sample. We limit the sample to planet radii to within $1-20R_\Earth$ (excluding 243 additional candidates). Brown dwarfs exist within this radius range, but secondary eclipse analyses within the \Kepler and \Ktwo automated software remove many of these emitting candidates. Furthermore, short-period brown dwarf occurrence is less than $6\times$ that of giant planets \citep{csi15}, rendering any remaining contamination negligible. After these cuts, we were left with 1,718 \Kepler planet candidates. Making equivalent cuts in the Scaling \Ktwo planet catalog, we retain 320 \Ktwo planets.

The two planet samples are displayed in Figure \ref{fig:planetcomp}. The population of super-Earths and sub-Neptunes, spanning $1-4R_\Earth$, appears divided by a dearth of planets near $2R_\Earth$, known colloquially as the radius valley \citep{ful17}. This sparsely populated region of parameter space highlights an  evolutionary process, where planets with thick H/He atmospheres and low surface gravity undergo mass loss (i.e., photoevaporation: \citealt{owe17}; core-powered mass loss: \citealt{gup18}), removing much of this thick atmosphere. These stellar proximity dependent mechanisms separate the sub-Neptunes from the super-Earths. We chose to use a physically motivated bound for these planet classes and employed the empirically derived radius valley period function in \citet{ho23}\footnote{We used the two-dimensional support-vector machine model from \citet{ho23}, which was derived using the \Kepler sample alone. Thus, we assumed the valley would be consistent within the \Ktwo sample, as shown by \citet{har20}. However, we acknowledge that differences may arise from dissimilarities in the underlying stellar mass populations, but expect these variations to be minimal.} as our divider. The separation between the other classes of planets is less clear; thus, we used $4-8R_\Earth$ for sub-Saturns and $8-20R_\Earth$ for Jupiters.

\section{Forward-Modeling with {\tt ExoMult}}
\label{sec:FM}
The aim of this paper is to compare \Kepler and \Ktwo planets ---when possible--- and to expand the stellar host parameter space of the \Kepler planet sample to refine existing trends. In order to make these two planet catalogs compatible, it is essential to properly account for  each sample's unique completeness features. To do so, we implemented our forward-modeling software {\tt ExoMult} \citep{zin19,zin20b}, which simulates the population of observed planets given some initial parent distribution. forward-modeling provides a straightforward method of accounting for catalog differences and sample reliability. This software package draws planets from around a sample of \Kepler and \Ktwo stars, subjecting each simulated planetary system to the geometric and instrumental selection criterion for each corresponding mission. Within {\tt ExoMult} we used the \cite{hsu19} and \cite{zin21} completeness functions to respectively account for \Kepler and \Ktwo detection/vetting biases. To simplify our calculation, we did not attempt to match the cataloged system multiplicities, treating each planet as an independent detection. If multiplicity was required to reproduce the \Kepler and \Ktwo observed planet total, we drew all multiplanet systems from a perfectly aligned disk (i.e., no mutual inclination). We also assumed all planets have zero eccentricity, which is motivated by the low-eccentricity population distributions found for these short-period planets that are sensitive to tidal circularization (e.g., \citealt{sha16}). Some warm giants considered in this study have heightened eccentricities, but our assumption can be made without loss of generality due to competing completeness effects. High-eccentricity transits on average project a shorter transit duration and orbit closer to the host star, reducing the signal strength while increasing the transit probability. Miraculously, these effects nearly cancel out \citep{bur08}, with some minor complications in multiplanet systems (see Section 8.5 of \citealt{zin19}).

\subsection{Model}
 The simulated planets in this study were initially drawn from a broken power law $g(P)$ in orbital period space and a single power law $q(R)$ in planet radius space as has been done in numerous other exoplanet occurrence studies \citep[e.g.,][]{you11,pet13b,muld18,zin19}.
 
\begin{equation} \label{eq:occper}
\begin{split}
g(P) &\propto
\begin{cases}
P^{\beta_1} & \text{if $P<P_{br}$} \\
P^{\beta_2} & \text{if $P \geq P_{br}$} \\
\end{cases} \\
q(R) &\propto  R^{\alpha} ,
\end{split}
\end{equation}
where $P_{br}$ represents the corresponding break in the period power law and the $\beta$ values are the scaling model parameters. Inherent in this model is the assumption of radius and period independence. Several studies \citep[e.g.,][]{ful17,van18} have identified a population valley that separates the super-Earths from the sub-Neptunes, indicating the existence of covariance between these parameters. By partitioning the planet sample along this valley, we minimize such complications. Our occurrence model $n$ was then normalized by a factor $f$, which represents the number of planets per star within the range of our sample:

\begin{equation} \label{eq:occ}
\frac{ d^2n}{d \log P \: d \log R} = f\: g(P)\: q(R).
\end{equation}
 
\subsection{Optimization}

To optimize this forward-modeling procedure against our observed planet samples, we relied on the methodology presented in \cite{zin20b}. In brief, the cumulative distribution function (CDF) for the simulated population was compared to the observed population for each relevant parameter. The Anderson-Darling test statistic, which provides an exponential scaling with the likelihood function, was used to identify more favorable models. This procedure is effective in replicating the shape of the underlying population, but CDFs are sample size independent, providing no constraint on the total number of planets. To constrain the number of planets expected, we relied on a modified Poisson likelihood function (see Equation 3 of \citealt{zin20b}), which provides a probabilistic comparison of the normalization of two drawn populations.

The posterior distribution for each corresponding parameter was measured using a Bayesian framework. We assumed uniform priors for all corresponding parameters with physically motivated bounds (i.e., we did not allow for negative occurrence fractions). Overall, we found that all imposed bounds exceeded the $10\sigma$ range of any identified model parameters, indicating the posteriors were not impacted by these boundary choices. To measure the posteriors for each parameter, we used the {\tt emcee} affine-invariant MCMC algorithm \citep{for14}.

\section{Results}
\label{sec:Results}
Here we focus on how the \Kepler planet occurrence rates compare with the \Ktwo rates for the four relevant planet classes.

\begin{figure*}
\centering \includegraphics[width=\textwidth{}]{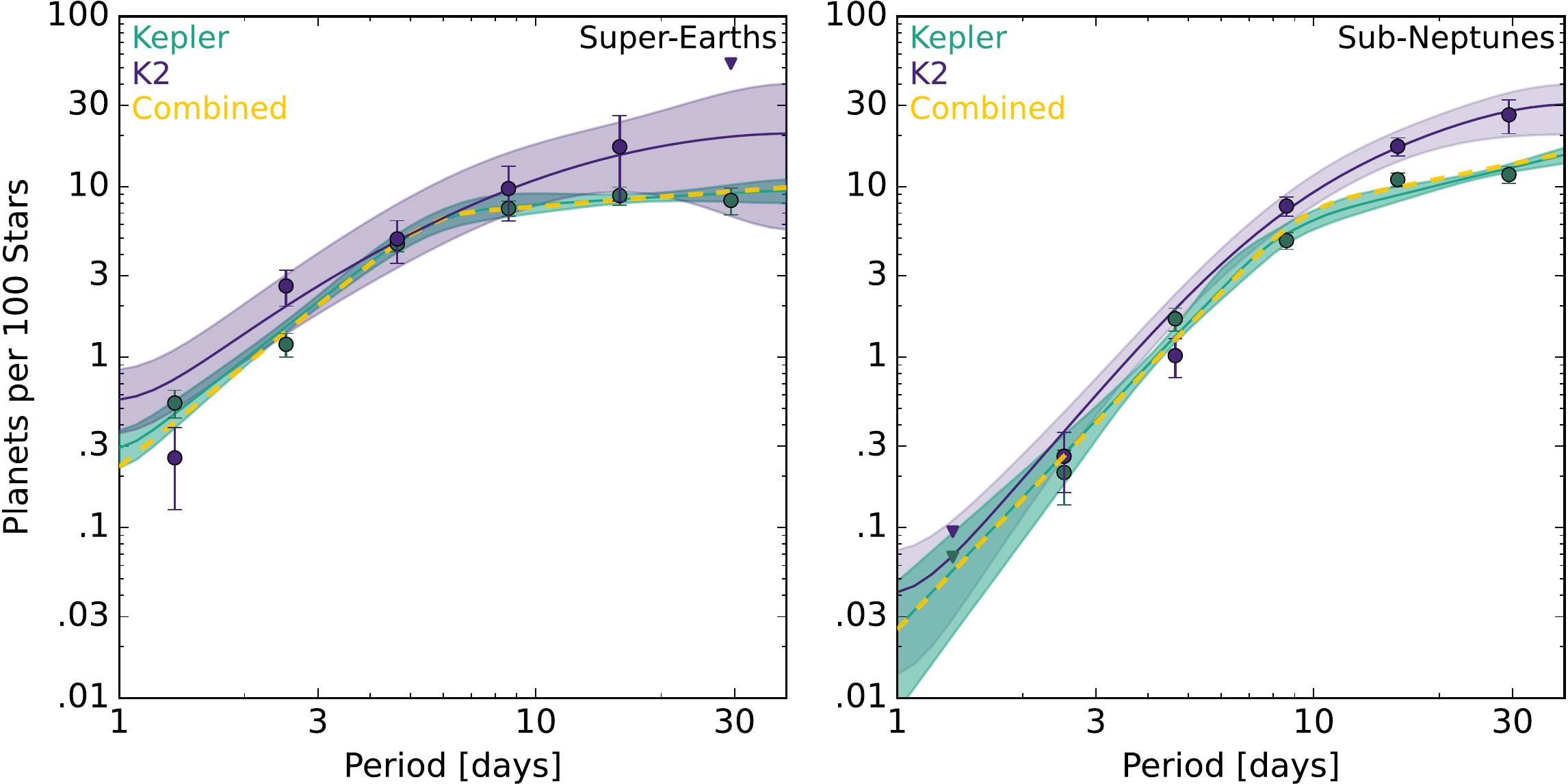}
\caption{Our output occurrence models for the super-Earths and sub-Neptunes. The shaded regions illustrate the $1\sigma$ range for each respective model. Over the top of our optimized models we provide binned occurrence values for reference. These binned values help guide the eye, but the fits were accomplished in the full unbinned data-set. The triangle markers are $3\sigma$ upper bounds, representing sparsely populated regions of parameter space. The gold dotted line shows the best-fit combined \Kepler and \Ktwo model. The parameter values for these models are provided in Table \ref{tab:compare}.
\label{fig:supsubCompare}}
\end{figure*}

\subsection{Super-Earths}
For the super-Earths we found increasing planet occurrence from 1 to 6 days for both \Kepler ($\beta_1=1.9\pm0.3$) and \Ktwo ($\beta_1=1.7\pm0.4$), which is shallower than numerous other occurrence calculations (i.e., \citealt{pet18Met} found $\beta_1=2.4\pm^{0.4}_{0.3}$). Previous studies have used a constant radius upper limit (usually around 2 $R_\earth$); however, we chose a functional form that follows the physical limits associated with the radius valley \citep{ful17}. This flattening of short-period occurrence is due to our empirically motivated assignment of super-Earth classification for a wider range of radii at short periods. Work by \cite{ber22}, who also used a nonfixed upper radius limit, found a comparable $1.44<\beta_1<2.90$ for this stellar mass range. This occurrence reduction within 6 days may be indicative of the inner edge of the gas disk \citep{mul15} having been truncated by the host star's magnetosphere \citep{lee17}, producing a distinct pile-up location at which planets may have an enhanced formation rate \citep{bol14} or may be preferentially trapped in the process of migration \citep{bai16}. Our noted reduction in $\beta_1$, from previous studies, indicates that super-Earth formation has a weak period dependence near the host star.

Around 6 days, planet occurrence turns over (\Kepler: $P_{br}=5.9\pm1.3$ days; \Ktwo: $P_{br}=8.9\pm^{7.4}_{6.24}$ days). Both \Kepler and \Ktwo $P_{br}$ values are statistically consistent with the 7--10 day value found by previous studies \citep{you11,how12,bur15,pet18Met}. However, a reduction in our \Kepler model ---which drew from the same planet sample as the previous works--- suggests a more nuanced interpretation. Furthermore, our combined model, which performed a joint analysis of \Kepler and \Ktwo populations, produced an even closer-in population peak ($P_{br}=5.6\pm1.1$ days). The root of the break reduction can again be traced to our upper limit for super-Earth classification. \citet{ber22}, who also used a nonfixed upper radius limit, found a turnover period more in alignment with our values ($6<P_{br}<12$ days) for a comparable mass range, using the Kepler sample alone. 

If planet cores were uniformly distributed in log-period space, $P_{br}$ is expected to shift with stellar mass or insolation flux, providing hints of the underlying mass loss mechanism. \citet{pet22} was unable to discern a clear trend in $P_{br}$ using a fixed $1.7R_\Earth$ boundary, partially attributing this obfuscation to a nonuniform core distribution. However, our flattening population model suggests a more uniform primordial core-period distribution, suggesting trends with $P_{br}$ may remain informative with careful planet classification. We leave further analysis of this trend with both \Kepler and \Ktwo for future studies. Regardless, it is clear that any negatively sloped functional radius bound will reduce $P_{br}$ by capturing more short-period planets and reducing the number of longer-period planets considered super-Earths, as noted in \cite{lop18}.

 For \Kepler we found $\beta_2=0.2\pm0.2$, which is consistent than many previous studies (i.e., \citealt{pet18Met}: $\beta_2=-0.3\pm0.2$). This trend remains flat even with our choice of a functional radius upper limit.  The removal of contaminating long-period sub-Neptunes from our super-Earth sample provides a minor correction. \cite{ber22} found a range of values ($-0.1<\beta_2<0.3$) using a nonconstant radius limit, aligning with our \Kepler model parameters. Our \Ktwo model is poorly constrained for the longer-period super-Earths ($\beta_2=-0.5\pm1.2$). Only five \Ktwo planets in our sample exist beyond 10 days. Overall, the \Ktwo occurrence model is within $1\sigma$ of our \Kepler model at all periods, suggesting consistent planet abundances. 

To provide a more refined population model for super-Earths, we combine both \Kepler and \Ktwo samples and provide the optimized model parameters in Table \ref{tab:compare}. Overall, \Kepler's abundant sample dominates the combined model fitting. However, the additional \Ktwo planets allow us to provide a more precise occurrence model for the local galactic super-Earth population.

\begin{deluxetable*}{lrccccc}
\tablecaption{The model parameters for our optimization of \Kepler and \Ktwo planet occurrence. The models are separated into planet classes: super-Earths (SE), sub-Neptunes (SN), sub-Saturns (SS), and Jupiters (J).   \label{tab:compare}}
\tablehead{\colhead{Class} & \colhead{Mission}& \colhead{$f$} & \colhead{$\alpha$} & \colhead{$\beta_1$} & \colhead{$P_{br}$} & \colhead{$\beta_2$}}
\startdata
SE &  &   &  &  & &    \\
 & \Kepler & $0.31 \pm 0.02$ & $-1.0\pm0.2$ & $1.9\pm0.3$ & $5.9\pm1.2$ & $0.2\pm0.2$ \\
 & \Ktwo & $0.48\pm0.21$ & $-1.9\pm0.7$ & $1.7\pm0.4$ & $8.9\pm^{7.4}_{6.24}$ & $0.5\pm1.2$ \\
 & Combined & $ 0.31\pm 0.02$ &$ -1.1\pm 0.2$ & $2.0\pm0.2$ & $5.6\pm1.1$ & $0.2\pm0.2$ \\
SN &  &   &  &  & & \\
& \Kepler & $0.28 \pm 0.01$ & $-1.5\pm0.1$ & $2.6\pm0.5$ & $8.5\pm2.4$ & $0.6\pm0.2$\\
 & \Ktwo & $0.59\pm0.10$ & $-2.0\pm0.3$ & $2.8\pm0.6$ & $10.9\pm^{5.0}_{3.9}$ & $0.9\pm0.7$ \\
 & Combined & $ 0.30\pm 0.01$ &$ -1.7\pm 0.1$ & $2.5\pm0.4$ & $9.5\pm2.0$ & $0.5\pm0.2$ \\
SS &  &   &  &  & &  \\
 & Combined & $ 0.019\pm 0.002$ &$ -2.7\pm 0.6$ & $1.7\pm0.5$ & $8.7\pm^{7.8}_{4.8}$ & $0.5\pm0.5$ \\
J &  &   &  &  & &  \\
 & Combined & $ 0.011\pm 0.002$ &$ -0.9\pm 0.4$ & $2.7\pm1.7$ & $3.4\pm^{1.9}_{0.8}$ & $-0.1\pm0.3$ \\
\enddata
\end{deluxetable*}

\subsection{Sub-Neptunes}

Sub-Neptunes have low densities due to their thick H/He atmospheres, which are susceptible to mass loss through either early-stage integrated XUV flux photoevaporation \citep{owe17} or residual heat from within the planet's core \citep{gup18}. This transitional planet class is uniquely positioned to test and bound the formation processes responsible for gas-giants and terrestrial planets. Our lower radius limits for sub-Neptunes is motivated by an observed dearth of planets expected from these mass loss mechanisms.

Previous studies have found occurrences of around 30\% for both sub-Neptunes and super-Earths orbiting FGK stars within 100 day periods \citep{you11,how12,bur15}, indicating an interconnected formation pathway. Since a vast majority of these planets orbit at shorter periods, our results confirm this consistency, finding $\sim30\%$ occurrence within 40 days for both populations. This population similarity may provide further evidence of a common origin for these two populations. We provide further discussion of this topic in Section \ref{sec:subsup}.    

An additional population deficit has been identified for sub-Neptunes with periods less than 10 days \citep{bea13,maz16}, where comparable period super-Earths are more abundant. This region of parameter space is  well sampled within the \Kepler mission catalog; thus, some physical process must be at play. The sub-Neptune desert likely reflects the saturation of photoevaporation or core-powered mass loss, removing the envelopes of nearly the entire planet class radius range and generating an abundant population of short-period super-Earths with thin H/He atmospheres. Here, we found a consistent trend in both the \Kepler ($\beta_1=2.5\pm0.5$) and \Ktwo ($\beta_1=2.8\pm0.6$) sub-Neptune populations, which is in agreement with previous studies that used a constant radius bound (i.e., \citealt{pet18Met}: $\beta_1=2.3\pm0.2$). This consistency and minor trend enhancement is surprising. Naively, it might seem that a functional radius bound would produce a steeper trend by eliminating super-Earth contaminants, but \citet{pet18Met} used a more limited sample of hosts ($4700<T_{\mathrm{eff}}<6500$ K). Since planetary mass loss mechanisms have stellar mass dependencies, it is likely our trend dictates a purer planet classification, but is convolved with a wider range of stellar masses, blurring the overall short-period trend.

We observed a break in occurrence at $P_{br}=8.5\pm2.4$ days for \Kepler and $P_{br}=10.9^{5.0}_{3.9}$ days for \Ktwo. These trends are within parameter uncertainty and agree with previous works \citep{you11,how12,bur15}, who found $P_{br}$ around 10 days. Despite their statistical similarity, Figure \ref{fig:supsubCompare} depicts an overall occurrence deviation between the \Kepler and \Ktwo sub-Neptune population near 10 days. \citet{pet22} noted a similar normalization increase when comparing differences in host mass within the \Kepler sample. Less-massive stars tend to harbor more sub-Neptunes, while maintaining a comparable trend in period-space. Our \Ktwo stellar sample represents a slightly less-massive population when compared to the \Kepler sample ($\Delta \tilde{M_\star}\sim-0.04M_\sun$). This underlying stellar population difference likely explains the observed offset near 10 days. However, the \Ktwo trend presented here does not provide a clear normalization offset and seems to enhance longer period planets more. \citet{pet18Met} found a similar normalization trend, suggesting metal-rich stars harbor more sub-40-day sub-Neptunes. In Section \ref{sec:metal} we show that \Ktwo stars are generally metal poor compared to the \Kepler population ($\Delta \tilde{\mathrm{[Fe/H]}}\sim-0.1$ dex), conflicting the observed shift in \Ktwo sub-Neptunes. The reality is likely a convolution of both stellar mass and [Fe/H]. 

Beyond 10 days, we observed a slight increase in planet occurrence (\Kepler: $\beta_2=0.6\pm0.2$ ; \Ktwo: $\beta_2=0.9\pm0.7$). This deviates from previous work, which found an occurrence plateau (\citealt{pet18Met}: $\beta_2\lesssim0$). This surplus of long-period planets can again be attributed to our function class limit, accurately capturing more sub-Neptunes at longer periods. 

Generally, we found consistency between the \Kepler and \Ktwo sub-Neptune samples, with a deviation for planets orbiting beyond 10 days. It is likely that these differences can be reconciled with careful accounting of the underlying stellar mass distribution. This apparent cohesion supports the population and formation trends established by the \Kepler planet catalog. For a more accurate occurrence model for the local galaxy, we provide a joint \emph{Kepler}/\Ktwo model in Table \ref{tab:compare}. The super-Earths and sub-Neptunes population models have remarkable similarity, with a notable divergence in the period break, providing information on the underlying mechanisms imprinted on the planet population. We provide a detailed discussion of this difference in Section \ref{sec:break}.

\subsection{Sub-Saturns}

Planet occurrence significantly drops for planets with $R>4R_\earth$ (e.g., \citealt{ful17}), indicating a unique formation pathway from that of the abundant sub-Neptune population. Empirically, these planets represent a population spanning $6-60M_{\earth}$ with little planet radius correlation, but a strong mass dependence on stellar [Fe/H] measurements \citep{pet17a}. This suggests that metal-rich disks manifest more-massive sub-Saturn cores. Using interior structure models (i.e., \citealt{lop13}) several planets have been identified with core mass fractions nearing 50\% (e.g., K2-19 b: \citealt{pet20} and TOI-257 b: \citealt{add21}). The prevalence of these heavy core sub-Saturns is at odds with the expectations of core accretion planet formation \citep{pol96}, which states that such massive cores should undergo runaway gas accretion and form Jovian sized planets.

\begin{figure*}
\centering \includegraphics[width=\textwidth{}]{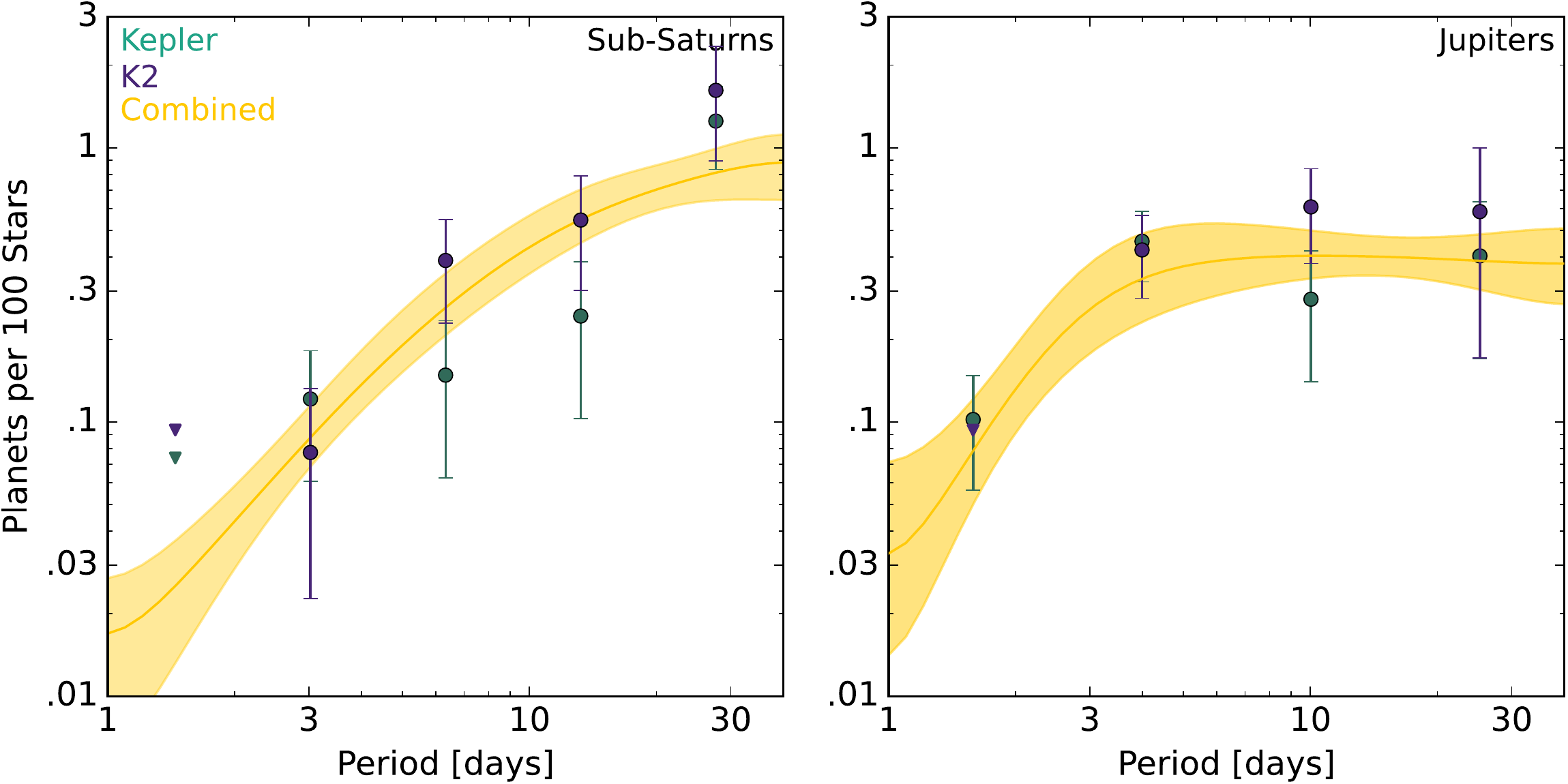}
\caption{The same output occurrence models as those in Figure \ref{fig:supsubCompare}, but for the sub-Saturn and Jupiter populations. Independent modeling of the \Kepler and \Ktwo samples was not possible due to the limited number of detected planets in these classes. Thus, we provide a model and $1\sigma$ region for the joint population. The parameter values for these models are provided in Table \ref{tab:compare}.
\label{fig:subjupCompare}}
\end{figure*}

Previous measurements of occurrence have found nearly 7\% of FGK stars host a sub-Saturn within 300 days \citep{pet18Met}. However, a majority of these planets lay at larger periods (>100 days). Adjusting existing results to the more limited 40-day period range of \Ktwo, they predict $\sim2\%$ occurrence. Consistently, we found that $1.9\pm0.2\%$ of FGK stars in the \Kepler and \Ktwo fields host sub-Saturns within 40-day periods (see Figure \ref{fig:subjupCompare}). The Jovian planets offer a remarkably comparable frequency ($1.1\pm0.2\%$, see Section \ref{sec:jov}).

Neither the \Kepler nor \Ktwo planet samples independently constrain our broken power law model, but visual inspection of the occurrence values in Figure \ref{fig:subjupCompare} show consistency between \Kepler and \Ktwo. Thus, we provide a joint analysis for a more robust population depiction. A steep decrease in occurrence is seen for planets with periods less than 10 days ($\beta_1=1.7\pm0.5$) matching previous results using the \Kepler sample alone \citep{bea13,maz16}. Unlike the sub-Neptune population, these more-massive planets are not significantly impacted by early-stage photoevaporative mass loss \citep{ion18}, requiring some additional mechanism like high-eccentricity migration in concert with tidal disruption \citep{owe18} to carve out the short-period region of parameter space. Alternatively, disk truncation may reduce in-situ formation in this material-barren region of parameter space. Beyond the period break at $8.7\pm^{7.8}_{4.8}$ days, sub-Saturn occurrence flattens ($\beta_2=0.5\pm0.5$). This significant population of warm sub-Saturns deviates from the Jovian population trend, suggesting a unique formation pathway. In Section \ref{sec:warmSaturns} we provide further remarks on potential formation scenarios for sub-Saturns.

Overall, we found consistency with the \Kepler and \Ktwo samples for the sub-Saturn population and provide a refined population trend, lending credence to a complex formation pathway for gas-giants.

\subsection{Jupiters}
\label{sec:jov}
The planets considered here are hot- and warm Jupiters, which straddle the 10 day period marker. Their formation mechanisms are the subject of significant debate. Three major theories have been proposed to explain hot-Jupiters, which seem to disobey standard core-accretion planet formation models. These gas-giants may be born at a larger orbital radius and driven to a high eccentricity orbit, where tidal forces exhaust orbital angular momentum (high-eccentricity migration: \citealt{wu03,fab07,nag08,bea12}). Alternatively, these planets may have migrated inward through interactions with the disk itself (disk migration: \citealt{gol80,lin96}). Tuned correctly, in-situ formation may also be possible for these hot-Jupiters \citep{bat16,boy16}. Warm-Jupiters, which span a range of 10-100 day periods, offer an origin further shrouded in mystery. Measurement of their host star's obliquity, appear to favor orbital spin-axis alignment, suggesting warm Jupiters have a more quiescent formation history than their hot counter parts \citep{mal22}.

 Like our sub-Saturn population, the \Kepler and \Ktwo samples are limited in their ability to independently constrain our model. Thus, we relied on visual inspection of Figure \ref{fig:subjupCompare} and observed occurrence consistency across missions. We provide a joint analysis for a more robust population analysis. Hot- and warm Jupiter occurrence rates appear to follow a broken power law, with the occurrence trend turning over around a three day orbital period. The sharp decrease within three days ($\beta_1=2.7\pm1.7$) may be the outcome of tidal disruption events.

 The three-day pile-up was first identified through early RV surveys (e.g., \citealt{udr03}), which observed this peak in short-period giant occurrence. The emergence of the early \Kepler sample brought into question the three-day pile-up, as transit demographic analyses failed to replicate the existing RV population feature \citep{how12,fre13}. \citet{daw13} later recovered a surplus of hot-Jupiters at three days around the metal-rich \Kepler stellar population, suggesting enhanced gas-giant production and a planet-planet scattering origin for some fraction of systems. \citet{san16} fully resolved a peak near 5 days in the \Kepler sample with careful removal of astrophysical false positives, reducing the tension with RV demographics. However, this two day offset between RV and transit surveys remained unsettled. A three day period corresponds to two times the stellar Roche radius, the expected limit of tidal disruption \citep{ras96,owe18}. If hot-Jupiters formed further out in the disk and attained heightened eccentricity through planet-planet scattering (or secular chaos \citealt{wu11}), tidal circularization would force these planets to migrate inward to short periods. This process is truncated by the tidal disruption radius, where orbital crossings tear the planet apart, leading to a population turn over around three days. We observed an occurrence break at $3.4\pm^{1.9}_{0.8}$ days for our combined \Kepler and \Ktwo Jupiter population, where the increased sample size shifts the break closer to the expected three-day limit. This finding relieves tension between RV and transit demographics, providing a robust measure of the three-day population break with transiting planets and therefore further evidence of high-eccentricity migration.

 The longer-period population, known as the period valley (e.g. \citealt{wit10}), is less clear in its origin. Many studies, including \citet{san16}, observed a dip near 10 days followed by increasing warm Jupiter occurrence out to 100 days. The inclusion of the \Ktwo sample smooths out this dip, yielding a flat distribution in $\log P$ out to 40 days ($\beta_2=-0.1\pm0.3$). This occurrence plateau is difficult to rectify with high-eccentricity migration, since it creates a binomial population of cold and hot-Jupiters (i.e., too few warm Jupiters; \citealt{wu11,petr14,petr16}).  Perhaps some combination of disk and high eccentricity migration blend to form the existing population of warm Jupiters \citep{daw18}. Our smoothing of the 3--40 day distribution with the joint \Kepler and \Ktwo sample further validates the complex origin of these planets.

It is important to note that metallicity is strongly correlated with hot-Jupiter occurrence \citep{fis05}; thus, these populations are significantly skewed towards metal-rich stars. Further discussion of the metallicity dependence is provided in Section \ref{sec:metal}.

\section{Trends in Stellar Host Parameters}
\label{sec:trends}
Here we consider known stellar trends in spectral class ($T_\mathrm{eff}$) and metallicity ([Fe/H]) and examine how the addition of \Ktwo planets modifies these parameters. The \Ktwo catalog only provides a 19\% increase in the total planet sample (from \Kepler alone). However, the Scaling \Ktwo sample provides a 28\% increase in bright hosts ($m_v<14$), where ground-based stellar characterization is readily feasible. Capitalizing on this fact, we can refine previous occurrence trends with the addition of \Ktwo planets. We also introduce a new trend that takes into account the host star's galactic oscillation amplitude ($Z_\textrm{max}$). 

In each subsection we discuss additional cuts made in the stellar and planet sample to ensure purity in the metallicity and galactic oscillation amplitude measurements. Large homogeneous surveys like \GAIA and LAMOST provide suffcient coverage of the parent stellar population, in that a large enough sample is available to accurate retain the full sample's parameter distributions, as shown in \citet{pet18Met}. However, our quality constraints may impact the overall occurrence normalization. To address this issue we introduce a correction factor ($\kappa$) to preserve the full catalog occurrence normalization. Further discussion and the derivation of this factor can been found in Appendix \ref{sec:norm}.  We use the following model to describe all three parameters:

\begin{equation} \label{eq:trend}
\frac{d^3n}{d T_\textrm{eff} \; d \textrm{[Fe/H]} \; d \log Z_\textrm{max}} \propto \kappa\: \cdot 10^{\textrm{[Fe/H]}\cdot\lambda +  \frac{T_\textrm{eff}}{1000K} \cdot\gamma} \: \cdot Z_\textrm{max}^\tau,
\end{equation}
where $\gamma$, $\lambda$, and $\tau$ are all tunable parameters. The metallicity term represents a power law accounting for [Fe/H] being a $log$ ratio of the element abundances, as done in previous [Fe/H] based occurrence calculations (i.e., \citealt{fis05}). We find a similar trend in \teff best replicates the observed population. This unique model may be due to some scaling with Planck's law, but the origin remains unclear. A power law was chosen for $Z_\textrm{max}$ out of simplicity and its ability to match the observed population trend. The model provided in Equation \ref{eq:trend} assumes independence between these three stellar parameters, which may be too simplistic. In Sections \ref{sec:metal} and \ref{sec:GalAmp} we will discuss these possible correlations and their impact on the trends. To optimize this model, we hold the best-fit combined (\Kepler and \emph{K2}) model parameters fixed for each respective planet class to minimize biases introduced by the underlying planet population. The results of our fitting are provided in Table \ref{tab:Trends}.

\begin{deluxetable*}{lcccccc}
\tablecaption{ Parameters for our combined \Kepler and \Ktwo stellar trend models. We provide the number of candidates in each respective sample in the two rightmost columns. \label{tab:Trends}}
\tablehead{\colhead{Class} & \colhead{Period (d)}  & \colhead{$\gamma$} & \colhead{$\lambda$} & \colhead{$\tau$} & \colhead{\emph{\Kepler}} & \colhead{\emph{\Ktwo}}} 
\startdata
SE &  &&&& \\
 & 1--40 & $-0.17\pm 0.02 $  & - & - & 802 & 51 \\
 & 1--10 & $-0.41 \pm 0.04$ & $0.5 \pm 0.1$ & - & 305 & 37  \\
 & 10--40 & $0.01 \pm 0.07$ & $0.0\pm 0.2$ & - & 133 & 4 \\
& 1--40 & $ -0.14 \pm 0.04 $ & - & $-0.30\pm0.06$  & 367 & 49 \\
SN &  &&&&  \\
& 1--40 &$ -0.25 \pm 0.02 $ & - & - & 797 & 219  \\
 & 1--10 &$ -0.42 \pm 0.05 $ & $ 1.2 \pm 0.1 $ & - & 160 & 68  \\
 & 10--40 & $-0.28 \pm 0.03$ & $0.26\pm 0.09$ & - & 307 & 104 \\
 & 1--40 & $ -0.25 \pm 0.03 $ & - & $-0.37\pm0.07$  & 322 & 190 \\
SS &  &&&&  \\
& 1--40 &$ -0.21 \pm 0.07 $  & - & - & 63 & 25 \\
 & 1--10 &$ -0.5 \pm 0.2 $ & $ 2.5 \pm 0.5 $ & - & 11 & 8  \\
  & 10--40 & $-0.23 \pm 0.15$ & $1.2\pm 0.4$ & - & 21 & 8 \\
J &  &&&&  \\
 & 1--40 & $ -0.01\pm 0.08 $ & - & - & 58 & 25  \\
 & 1--10 &$ 0.0\pm 0.2 $ & $ 2.6 \pm 0.5 $ & - & 26 & 10  \\
& 10--40 & $-0.2 \pm 0.03$ & $1.6\pm 0.8$ & - & 6 & 2 \\
\enddata
\end{deluxetable*}

\begin{figure*}
\centering \includegraphics[width=\textwidth{}]{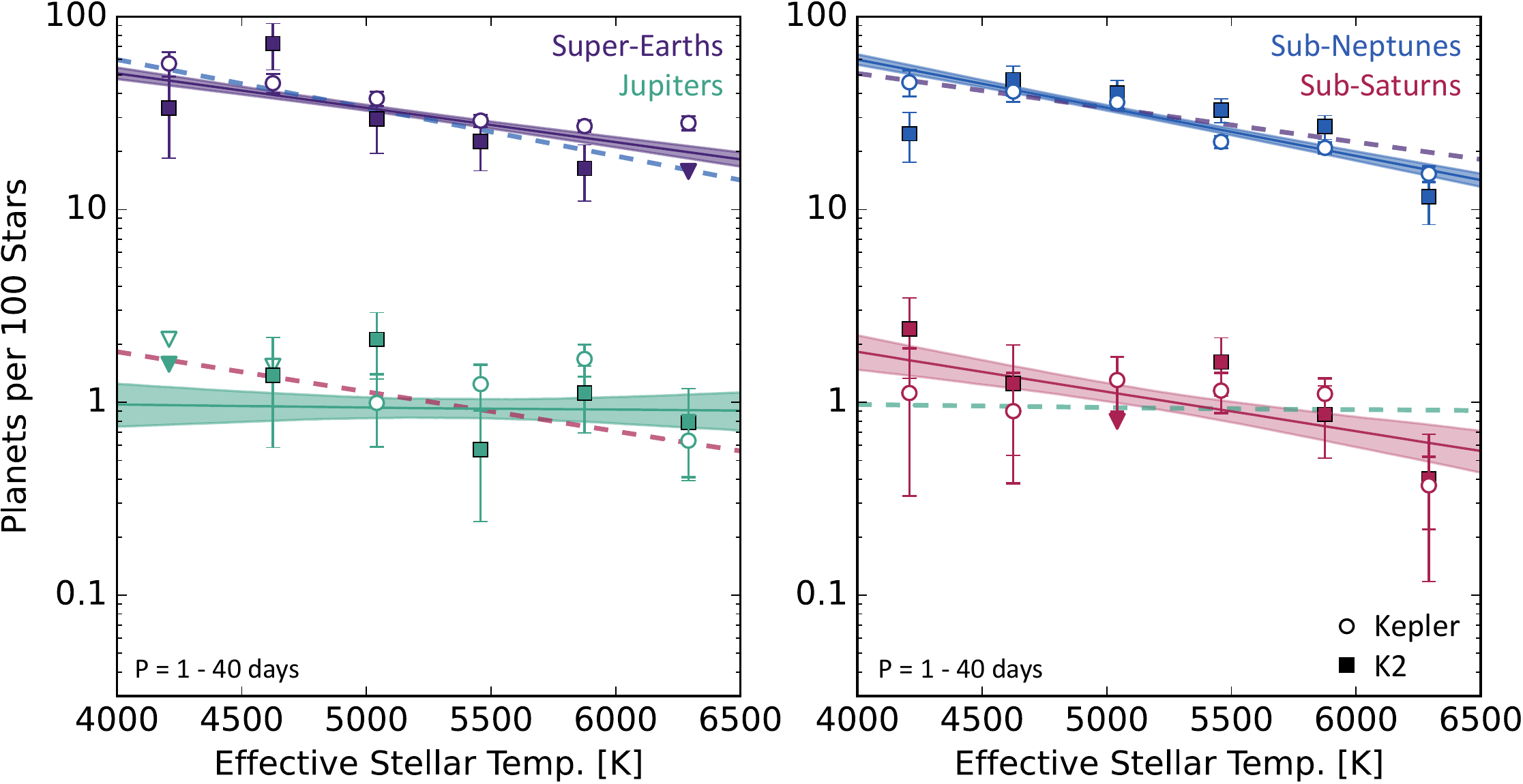}
\caption{ The occurrence for \Kepler (circles) and \Ktwo (squares) for each relevant class of planets as a function of effective stellar temperature, which is a proxy for spectral class and disk mass. The best-fit trend lines are displayed with the $1\sigma$ model regions shaded.  We included the best-fit trend of the adjacent plot as dotted lines for comparison. In this analysis, we set $\lambda$ and $\tau$ to zero to focus on \teff dependencies ($\gamma$) and to utilize the full population sample without loss of precision. The trend model parameters are provided in Table \ref{tab:Trends}. 
\label{fig:stellarCompare}}
\end{figure*}

\subsection{Spectral Class}
\label{sec:specClass}
More-massive stars are the product of a more-massive gas cloud contraction and should, correspondingly, host higher-mass protoplanetary disks early in their lifetimes. Observationally, this has been shown by \citet{and13}. Planet occurrence trends may provide insight into this natal state disk, as planets are born out of the stellar residuals. With RV detections, \citet{joh07} found giant planets are more common around massive stars, aligned with expectations that increased disk material inventory enables more planet formation. However, later occurrence analysis using transiting planets found fewer small (super-Earth and sub-Neptune) planets around more-massive stars \citep{how12,muld15}, indicating disk solids have a more complex relationship with these smaller planet classes. 

In Figure \ref{fig:stellarCompare} we investigate the impact of stellar spectral class on planet occurrence. Here, we use \teff as a proxy for stellar mass under the guise of established mass-temperature relationships \citep{kui38}. Many of our targets still lack the proper stellar characterization --via high-resolution spectra-- required for robust mass assessment; however, existing photometry allows for reasonable stellar temperature measurements ($\sigma_{\mathrm{Teff}}\sim100K$; \citealt{har20}). For this test we set $\lambda$ and $\tau$ to zero, enabling us to use the full catalog sample and avoid contamination from poorly constrained stellar characterization. For the remainder of the analysis performed in this subsection will we used the full stellar and planet sample.

Within our sample of FGK stars, we found decreasing occurrence for super-Earths ($\gamma=-0.18\pm0.02$) and sub-Neptunes ($\gamma=-0.25\pm0.02$) with increasing stellar temperature. These known trends are further amplified with the inclusion of \Ktwo planets, providing additional evidence for the anti-correlation of stellar mass and small planet occurrence. Both populations depict a nearly 75\% reduction in planet occurrence moving from a \teff of 4000K to 6500K, while the star mass increases nearly 75\% across these spectral classes. This counter-intuitive trend may be caused by a rise in outer giant formation, cutting off the flow of pebbles to the inner disk and reducing the material available for smaller planet formation \citep{mul21}.

\begin{figure*}
\centering \includegraphics[width=\textwidth{}]{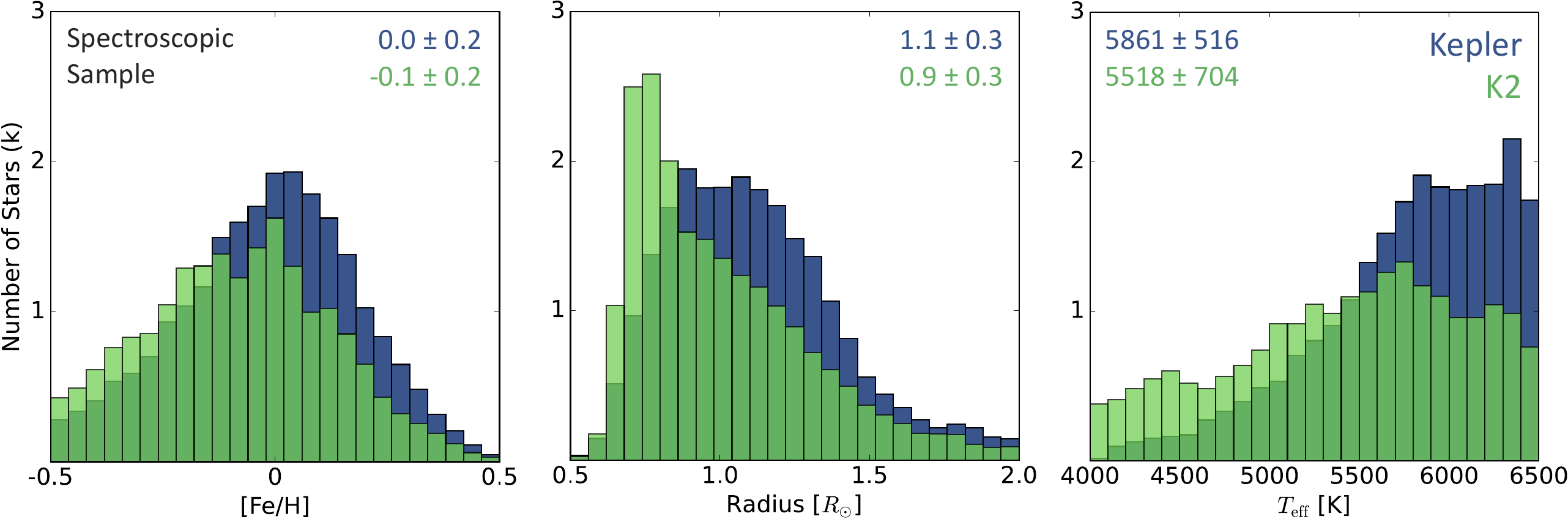}
\caption{ The distribution of our stellar sample with spectroscopically derived parameters. The corresponding distribution median and MAD values are provided in each parameter window.
\label{fig:spec_stellar}}
\end{figure*}

Stellar trends for transiting giants are difficult to resolve with the \Kepler sample alone, given the relatively narrow and peaked stellar parameter space probed by the Kepler host stars (see Figure \ref{fig:spec_stellar}). Using \Ktwo we can significantly improve the resolution of these trends due to the improved coverage of the edges of the stellar parameters probed across the two surveys. Intriguingly, we find decreasing sub-Saturns occurrence around more-massive stars ($\gamma=-0.21\pm0.07$). If the mass loss/disruption mechanism responsible for sculpting the edge of the sub-Saturn desert swells outward to longer periods for high mass stars, as suggested by \citet{hall22}, then such evolutionary processes may be responsible for the observed sub-Saturn occurrence reduction. 

We also observed a flat occurrence profile for Jupiters as a function of stellar spectral class ($\gamma=-0.01\pm0.07$), contrasting established RV results that found increasing occurrence as a function of stellar mass for M dwarfs \citep{joh07}. However, a more granular examination of our sample shows the population rate is increasing from 4000--4500K followed by a flattening, suggestive of a rise in occurrence from M to K dwarfs. We suspect a wider stellar sample would display Jupiter planet occurrence steeply increasing up to the early K dwarfs, at which point the occurrence rate flattens. Perhaps this turnover indicates some saturation point, where increasing the disk mass leads to no further hot- and warm Jupiter planet production. Using \tess, \citet{bel22} found evidence of decreasing hot-Jupiter occurrence around more-massive stars (AFG types), which may complicate this narrative; however, this trend remains tentative. We provide further discussion of the implications of this trend in Section \ref{sec:gasMetal}. Further analysis with direct measurements of stellar host mass will provide a more clear understanding of these apparent trends.

Within our sample there may exist degeneracies in \teff and [Fe/H]. Hotter and more-massive stars tend to be younger and have a higher metallicity. \citet{joh10} showed that a mass-metallicity plane could provide a more robust description of the observed gas-giant population. In Section \ref{sec:metal} we investigate the impact of composition on planet occurrence, while simultaneously fitting for a spectral class trend. In Section \ref{sec:GalAmp} we disentangle these two stellar features and resolve a galactic location dependence.

\subsection{Metallicity}
\label{sec:metal}
The natal disk inventory dictates the available building blocks for planet formation. By considering the current stellar abundances, as measured through spectral analysis, we are provided a glimpse into the composition of this natal disk. Early RV studies found strong correlations between Jovian planets and stellar iron (Fe) abundances (e.g., \citealt{fis05}), suggesting Fe is a key component in giant planet formation. Further analysis of RV planets found the trend weakened for smaller planets \citep{sou08,ghe10}, indicating some alternative relation with the disk iron content. This reduced correlation was corroborated with the \Kepler transiting sample \citep{pet18Met}, using HIRES and LAMOST spectra to measure the [Fe/H] for a subset of hosts and field stars.

We provide further refinement to this analysis by including planets from the \Ktwo sample. Up until this point we have been using a sample of stars that contain a mixture of spectroscopically and photometrically derived parameters, but the occurrence models used in Section \ref{sec:Results} only relied on quantities well constrained by photometry. [Fe/H] measurements are poorly constrained via photometry. Thus, to ensure a pure sample we removed targets without available spectra, leaving 22,985 \Kepler and 19,084 \Ktwo targets. Correspondingly, the planet sample was reduced to 969 \Kepler and 241 \Ktwo planets. From this pure sample \footnote{Our spectroscopic sample harbors metallicities with uncertainties ranging from 0.004 to 0.15 with a median uncertainty value of 0.03 dex. We did not account for uncertainty in our model optimization, but expect the impact to be minimal given the precision of our sample.} we can extract meaningful metallicity trends. It is important to note that this cut may systematically change the overall occurrence of any planet class. Thus an additional correction factor ($\kappa$) was introduced to preserve the initial class occurrence (see Appendix \ref{sec:norm} for more details). In optimizing our planet population models for this sample, we set $\tau$ to zero, allowing \teff and [Fe/H] dependence to be assessed without contamination from inaccurate galactic PM measurements. 

We previously noted that Equation \ref{eq:trend} assumes parameter independence, which may not be accurate. Thus, we examined the posterior chains of \teff and [Fe/H] and found little evidence of covariance among these parameters for any of the planet classes. We do not attempt to rule out any such correlations, but note that our assumption of independence was not wildly inaccurate, and provides a reasonable model for the associated stellar trends.

In Section \ref{sec:Results} we discussed various period breaks in each planet class, ranging from 3 to 10 days. These turnovers in occurrence seem to indicate some underlying physical processes, separating the populations. The exact origin of these population peaks remains unclear, but it seems plausible that either side of these breaks represents planets with unique formation histories. Although we find a range of $P_{br}$ values for each planet class, we chose to separate the planet population at 10 days to offer a direct comparison with the existing literature.

\begin{figure*}
\centering \includegraphics[width=\textwidth{}]{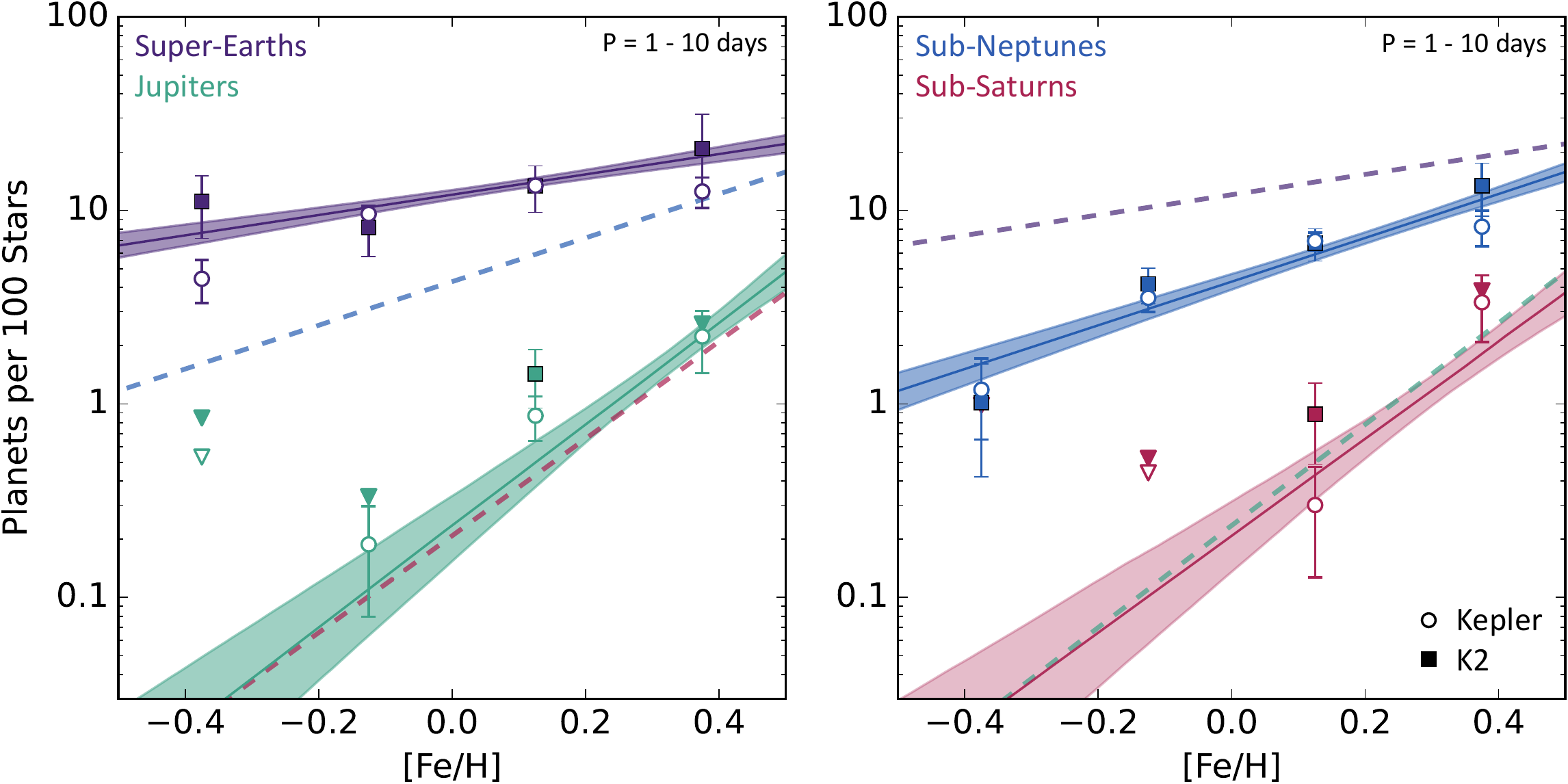}
\caption{The occurrence of short-period (1-10 day) \Kepler (circles) and \Ktwo (squares) planets as a function of [Fe/H]. The triangle shapes represent $3\sigma$ upper limits for the respective bin. The best-fit trend lines are displayed with $1\sigma$ model regions shaded. To highlight similarities between planet classes, we display the best-fit models of sub-Neptunes and sub-Saturns as dotted trends on the right panel. Likewise, we display super-Earth and Jupiter dotted trends on the left panel. The model parameters are provided in Table \ref{tab:Trends}. 
\label{fig:metalCompare}}
\end{figure*}

\subsubsection{Super-Earths and sub-Neptunes}

In Figure \ref{fig:metalCompare} we display sub-10 day planet occurrence as a function of stellar iron abundance from our combined \Kepler and \Ktwo sample. Super-Earths produce a trend ($\lambda=0.5\pm0.1$) consistent with the results of \citet{pet18Met} and \citet{wil22}, but with a 50\% reduction in uncertainty. These previous studies found a marginally significant trend ($\sim3\sigma$), but here we confirm to $5\sigma$ confidence that hot super-Earths have a positive metallicity dependence. We found similar agreement, with the previous works, for the population of hot sub-Neptunes ($\lambda=1.2\pm0.1$). These two planet classes have similar masses, but are differentiated by their outer envelope, suggesting an interconnected formation history. Photoevaporation models \citep{owe18} expect an increased mass loss in lower metallicity atmospheres, where cooling is less efficient. Hot sub-Neptunes are expected to undergo significant atmospheric removal due to the proximity to their host star, leaving behind a super-Earth with a thin H/He atmosphere. The predicted metallicity dependence is in alignment with our trends, which display a steep increase in the occurrence of hot sub-Neptunes around metal rich stars. Furthermore, we found a 70\% increase in the spectral class dependence (relative to the trend in Figure \ref{fig:spec_stellar}) in these short-period planets ($\gamma=-0.42\pm0.05$), indicating increased luminosity further reduces sub-Neptune occurrence, an expectation of photoevaporation models. If mass loss completely explained this occurrence trend, we might expect a similar-magnitude negative metallicity correlation for the hot super-Earth population. However, the observed positive occurrence slope for the hot super-Earths is likely due to increased core production at higher disk metallicities, an expectation of core accretion \citep{pol96}. It is clear that multiple mechanisms are at play within these populations, and further investigation is necessary to parse the magnitude of their effects.

\begin{figure*}
\centering \includegraphics[width=\textwidth{}]{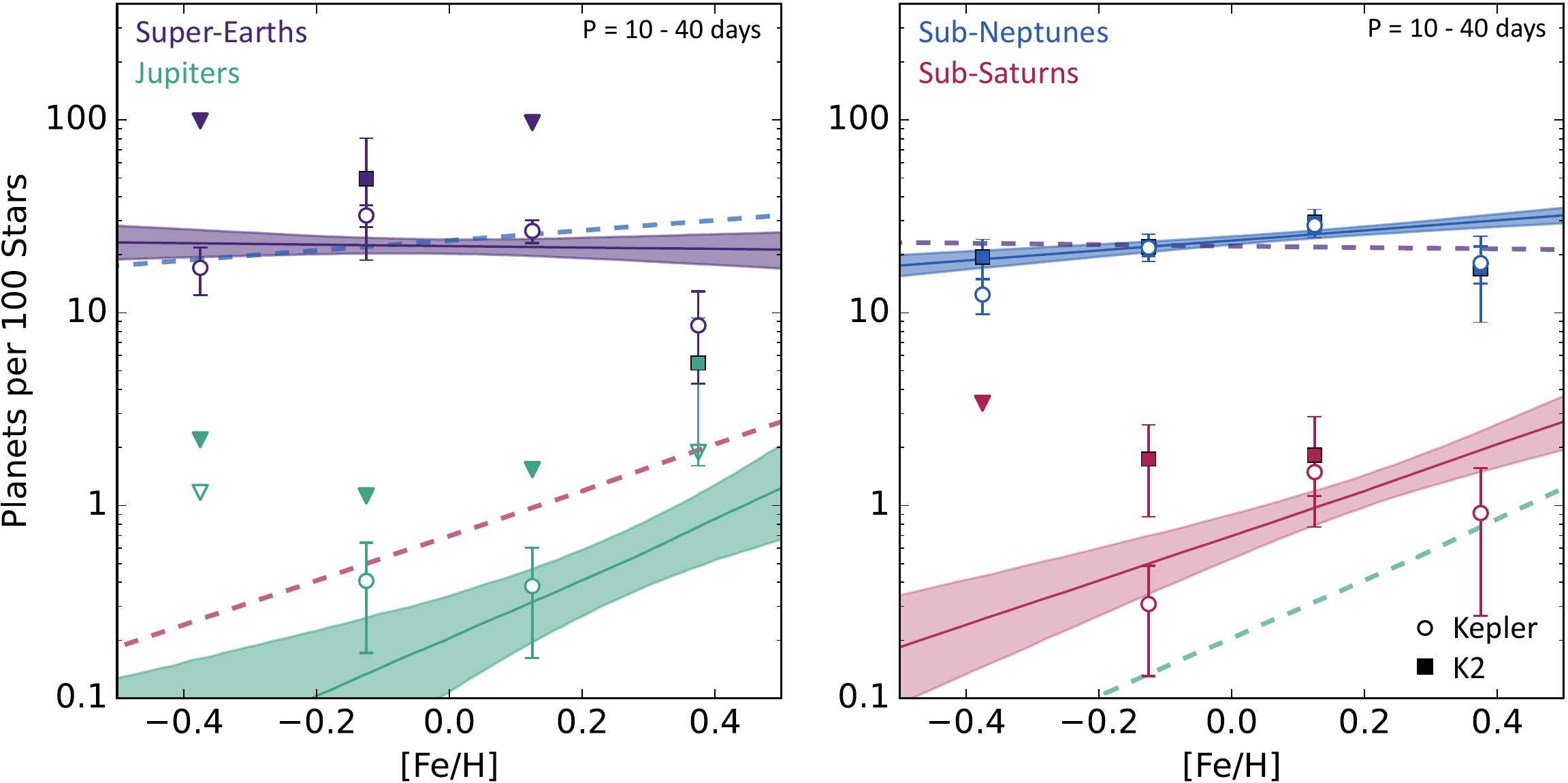}
\caption{ The occurrence of warm (10--40 day) \Kepler (circles) and \Ktwo (squares) planets as a function of [Fe/H]. The triangle shapes represent $3\sigma$ upper limits for the respective bin. The best-fit trend lines are displayed with $1\sigma$ model regions shaded. The trend values are provided in Table \ref{tab:Trends}. 
\label{fig:metalCompare_long}}
\end{figure*}

At longer periods (10--40 days; Figure \ref{fig:metalCompare_long}), the mass loss mechanisms carving out the radius valley define our population limits. Thus, occurrence trends in this region should highlight the natal formation of super-Earths and sub-Neptunes. We observed a negligible correlation with [Fe/H] ($\lambda=0.0\pm0.2$) and \teff ($\gamma=0.01\pm0.07$) for warm super-Earth occurrence. Sub-Neptunes present a slight correlation with stellar metallicity ($\lambda=0.26\pm0.09$) while maintaining a significant spectral class dependence ($\gamma=-0.28\pm0.03$). These trends are in agreement with previous studies \citep{pet18Met,wil22}, who measured [Fe/H] effects using a wider period range (10--100 days). Again, we reduce the uncertainties by $\sim 50\%$, further flattening the warm super-Earth trend. If standard core accretion is responsible for these correlations, we would expect more-massive cores around metal-rich stars, increasing sub-Neptune production while reducing super-Earth occurrence. Therefore, the flattening of the super-Earth trend provides tension here. Perhaps metal-driven cooling effects saturate at longer periods and these super-Earths represent a population of planets born with thin H/He atmospheres. \citet{rog21} found that $\sim20\%$ of the super-Earth population formed with a thin envelope, contrasting the primary expectation of an atmospheric mass loss origin. Furthermore, the isothermal cooling limits for super-Earth sized cores ($1-2M_\Earth$) provide a ceiling on the natal gas accretion, predicting the existence of a robust population of longer-period primordial super-Earths, where mass loss processes are impotent \citep{lee21}. Intriguingly, our models suggest warm super-Earth cores are formed independently of the stellar host star parameters, while sub-Neptune production decreases for more-massive early-type stars. In Section \ref{sec:break} we discuss this difference and the implications for formation. Close visual inspection of the warm super-Earth and sub-Neptune occurrence values suggest a broken power law may be more appropriate in modeling their metallicity dependence, further complicating their underlying genesis. We leave such analysis for future studies.

\subsubsection{Gas-Giant Metallicity Trends}
\label{sec:gasMetal}
Our sample of \Ktwo sub-Saturns and Jupiters provide an additional 16 and 12 planets, respectively. Since the existing \Kepler population only consists of 32 sub-Saturns and 32 Jupiters, this significantly increases the available sample. Compared to the smaller planet classes, we find a stronger metallicity trend for the gaseous hot sub-Saturn ($\lambda=2.5\pm0.5$) and hot Jupiter populations ($\lambda=2.5\pm0.4$), in agreement with previous \Kepler based work \citep{buch12,pet18Met,wil22}. Our reduced uncertainties bring these two population trends into alignment, indicating a similar disk metallicity dependence. Furthermore, we find a hot Jupiter correlation in agreement with previous RV surveys of local solar neighborhood ($\lambda=2.1\pm0.7$; \citealt{guo17}), strengthening this result.

Our 10--40 day populations indicate a consistent metallicity dependence for the warm sub-Saturns ($\lambda=1.2\pm0.4$) and the warm Jupiters ($\lambda=1.6\pm0.8$). We observed a trend reduction when compared to their short-period counterparts, and witness alignment with long-period RV trends (\citealt{joh10}$: \lambda=1.2\pm0.2$). Beyond 10 days, it appears sub-Saturns are $3\times$ more likely to occur than Jupiters. This is in conflict with standard formation models \citep{pol96}, which expect runaway accretion to take over in the sub-Saturn core mass range ($\sim10M_\Earth$). mass loss mechanisms like photoevaporation \citep{hall22} should have a greater impact on the short-period population, but we find this excess is unique to planets orbiting beyond 10 days, suggesting some primordial formation effects. Further discussion of this offset is provided in Section \ref{sec:warmSaturns}. 

It may be that the disk itself is throttling the production of more-massive warm Jupiters, but we find no evidence of any spectral class dependence ---which provides a proxy for stellar and disk mass--- when simultaneously fitting for metallicity ($\gamma=-0.2\pm0.3$). It seems that metallicity is the driving mechanism in the production of warm Jupiters and the disk mass is not a major limiting factor. Understanding the mechanism for this surplus will provide significant insight into sub-Saturn planet formation.

Overall, we find similar results with existing planet metallicity trends, but our inclusion of \Ktwo planets shows convergence in the trends for giants and smaller planets respectively.

\begin{figure*}
\centering \includegraphics[width=\textwidth{}]{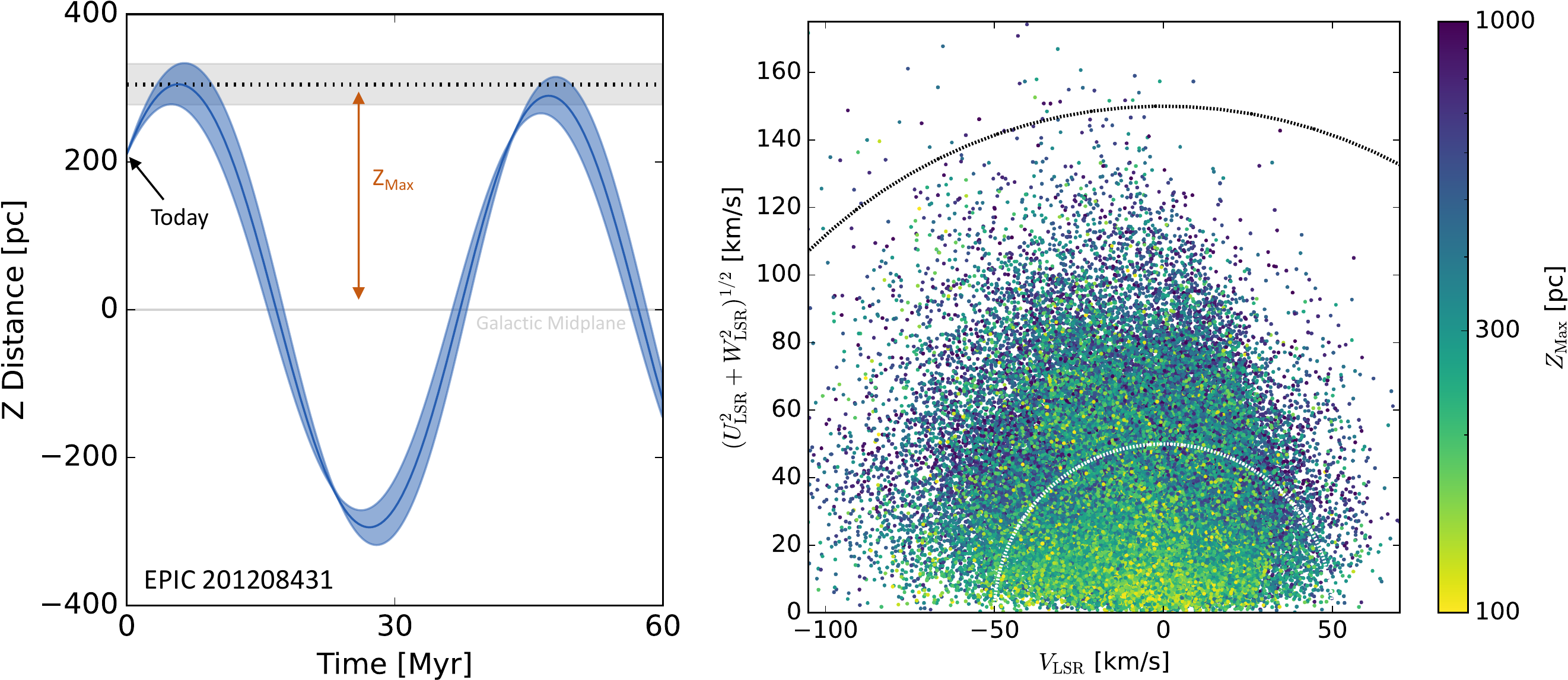}
\caption{\textbf{Left} An example of our galactic oscillation simulation for EPIC 201208431, showing how we derived our stellar amplitude values. Using the \texttt{Gala} software, in conjunction with the \GAIA DR3 proper motion and radial velocity parameters, we simulated each star's oscillation (dark blue line) about the galactic mid-plane, measuring the star's amplitude ($\textrm{Z}_\textrm{Max}$). We then repeated this N-body simulation 100 times, varying the \GAIA values within their uncertainty ranges. The oscillation $1\sigma$ range is displayed in light blue and the uncertainty in $\textrm{Z}_\textrm{Max}$ is shown in light gray. \textbf{Right} displays the corresponding Toomre diagram for our astrometric stellar sample. U,V, and W velocities were calculated in the local standard of rest frame (LSR). The 50 and 150 km/s total relative velocities have been provided as loose bounds for the thin and thick disks respectively. In general our integrated orbital $Z_{\textrm{Max}}$ values map on the this diagram as expected; low $Z_{\textrm{Max}}$, thin disk stars, have a much lower total relative velocity than high $Z_{\textrm{Max}}$, thick disk, counterparts.
\label{fig:amplitude}}
\end{figure*}

\subsection{Galactic Oscillation Amplitude}
\label{sec:GalAmp}
Galactic location remains of interest in planet formation given the unique element inventory of the differing galactic sub-structures. Thick disk stars have reduced iron and enhanced alpha element abundances, a product of Type II supernova dominance during the formation of these older stars \citep{wal62}. Furthermore, halo stars are a mixture of very metal poor (older) stars diluted by a population of alpha-enhanced (younger) stars, likely the outcome of young satellite population capture \citep{ven04}. Radially, a negative metallicity trend has been identified (e.g., \citealt{che12}), indicating stars near the galactic center are metal-rich compared to stars in the local solar neighborhood. These galactic sub-structures also harbor unique birth environments that may dynamically interact with planetary systems. For example, many halo stars were born in dense globular clusters \citep{har76} where an abundance of stellar interactions can produce complex planetary systems (i.e., \citealt{spa14}).

The current planet population is the outcome of natal disk composition and a history of dynamic interactions. Since stars are not born in isolation, but rather part of a dynamic galaxy, their location may provide a fossil record of planet formation throughout the history of the Milky Way, unveiling long-term formation processes and pathways not seen in younger planet populations. \citet{mct19} examined spatial planet occurrence of \Kepler hosts by considering \GAIA DR2 galactocentric velocities, finding that all differences between the host and non-hosting populations could be explained by selection effects. \citet{chen21} used LAMOST spectra and found a slight decrease in system multiplicity for thick disk hosts, suggesting an increased rate of instabilities for these older systems. Using \emph{TESS}, \citet{kol21} and \citet{bol21} constrained the hot Jupiter population in the metal-barren halo to less than $0.18\%$, predicting a minimum [Fe/H] formation threshold between -0.7 and -0.6 dex. Using the combined \Kepler and \Ktwo sample, along with proper motion measurements from \GAIA DR3 \citep{gaiadr3}, we examined our planet sample for trends in stellar location.

\begin{figure*}
\centering \includegraphics[width=\textwidth{}]{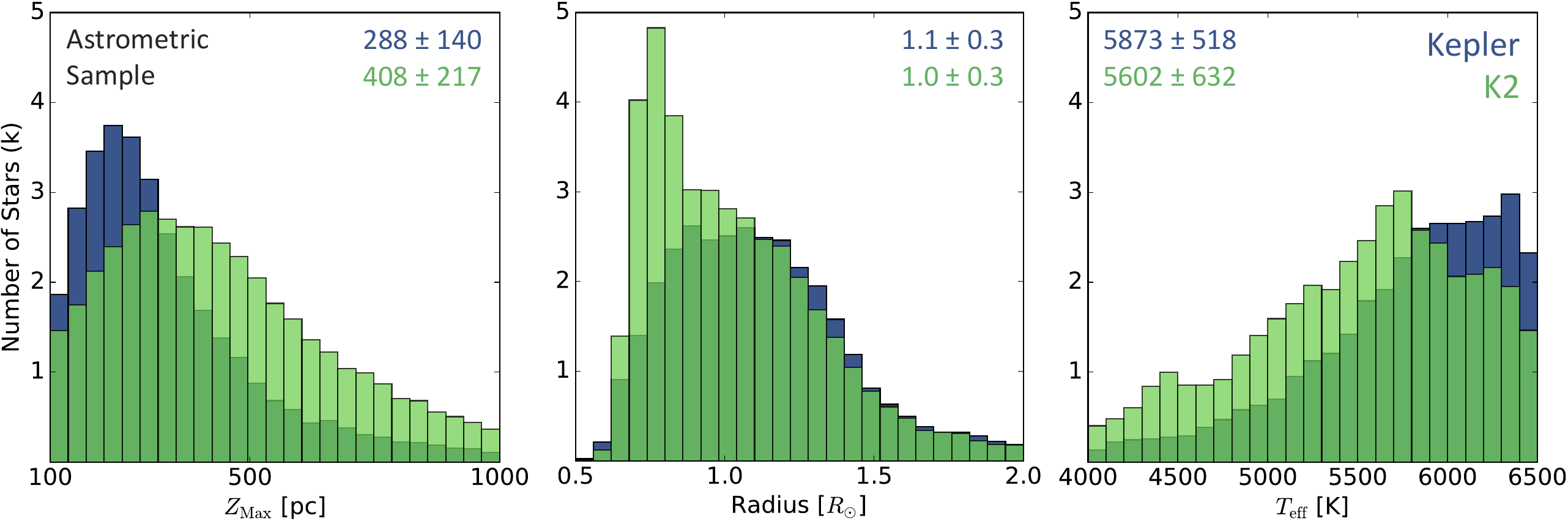}
\caption{The distribution of our stellar sample with precise astrometric parameters available. The corresponding distribution median and MAD values are provided in each parameter window. 
\label{fig:astro_stellar}}
\end{figure*}

A star's current position, relative to the galactic disk, can be determined by sky coordinates and parallax distance. However, these distances are merely a snapshot of their oscillation trajectory about the disk mid-plane. The amplitude of this oscillation is the characteristic that dictates the sub-structure membership (thin disk vs. thick disk).  Previous work has used kinematic properties to assign membership probability \citep{chen21}, binning the stellar population into separate sub-structures. This method is robust and attainable for nearly all stars in the \Kepler sample. However, it fails to capture subtle trends that may be convoluted due to dynamic diffusion and kinematic mixing of these discrete populations \citep{rix13}.We chose to consider a direct measure of the fully integrated orbital oscillation amplitude, using \GAIA DR3 proper motion (PM) and radial velocity (RV) measurements \citep{katz22}. To ensure consistency, we required all targets in this section to have measured \GAIA RVs and PM measurements. Implementing the \texttt{Gala} software \citep{gala} to model the stellar orbits about the galactic disk, we used a simple mass model for the Milky Way (derived in \citealt{bov15}). All stars then underwent 100 simulated oscillations about the disk, varying the PM and RV values according to the stated uncertainties to achieve a measure of the amplitude precision. We then removed all targets with amplitude uncertainties greater than 10\% to maintain purity (see Figure \ref{fig:amplitude}). Furthermore, we focused on targets with semi-amplitudes greater than 100 pc and less 1000 pc since this range contains a majority of planet sample and emphasizes the transition from thin to thick disk stars. The parameter distribution of this astrometrically pure sample is provided in Figure \ref{fig:astro_stellar} and a corresponding Toomre velocity diagram of our astrometric stellar sample is shown in Figure \ref{fig:amplitude}. This astrometric filtering removes 83,304 targets and 1,075 planets from our sample. As previously mentioned, these cuts may impact the overall occurrence normalization; thus, a correction factor was implemented (see Appendix \ref{sec:norm} for details).

In Figure \ref{fig:galTrend} we present the trend for $Z_\textrm{Max}$ as a function of planet occurrence for the super-Earth and sub-Neptune host populations. Overall, we found reduced planet occurrence at higher galactic amplitudes. As pointed out in \cite{mct19}, such galactic trends may be the outcome of sample selection effects. To address this issue, we simultaneously fit for \teff and observed that this trend remains significant ---super-Earths: $\tau=-0.30\pm0.06$; sub-Neptunes: $\tau=-0.36\pm0.07$. It is expected that stellar metallicity will decrease moving up in amplitude, naturally reducing planet occurrence. However, limiting the sample to stars with spectroscopic metallicities and precise astrometry reduces the planet sample to a non-statistical level. Instead of a direct simultaneous fit with stellar [Fe/H], we measured the expected metallicity-amplitude trend in the overlapping spectroscopic and astrometric stellar samples, finding a slope of $-0.275$ dex/kpc. \citet{Schl14} carried out a large survey of galactic chemistry gradients and found a weaker $-0.243$ dex/kpc trend ---indicating our derived gradient is likely over-representing the effect of metallicity. We then computed the expected occurrence-metallicity correlation for the full 1--40 day period range, yielding $\lambda$ values of $0.24 \pm 0.06$ for the super-Earths and $0.34 \pm 0.05$ for the sub-Neptunes. Putting these trends together, we conservatively estimated a super-Earth and sub-Neptune population reduction of $14\pm3\%$ and $19\pm2\%$ respectively over the first kpc above the galactic plane due to metallicity alone. This is significantly less than the $50\pm8\%$ and $56\pm7\%$ occurrence drops seen for the super-Earth and sub-Neptune populations. In other words, we find a greater than $4\sigma$ difference between our conservative metallicity estimate and the observed trends, indicating an alternative mechanism must account for the lack of high stellar amplitude planets. It is important to note that we forced $\lambda$ to be zero for this optimization, and that a more thorough accounting of the correlations between [Fe/H] and $Z_\textrm{Max}$, with a well parameterized stellar sample, may yield a reduced $\tau$ value. However, our quantification of the expected [Fe/H] contributions shows that the existence of even a strong correlation between location and metallicity will not rectify the detected $Z_\textrm{Max}$ trend. The physical amplitude is likely a proxy for some other parameter, like stellar age, but it is clear that [Fe/H] and \teff trends are not sufficient in replicating the observed occurrence rate gradient. In Section \ref{sec:gaxForm} we provide further remarks on the potential origin of this galactic trend.

\begin{figure*}
\centering \includegraphics[width=\textwidth]{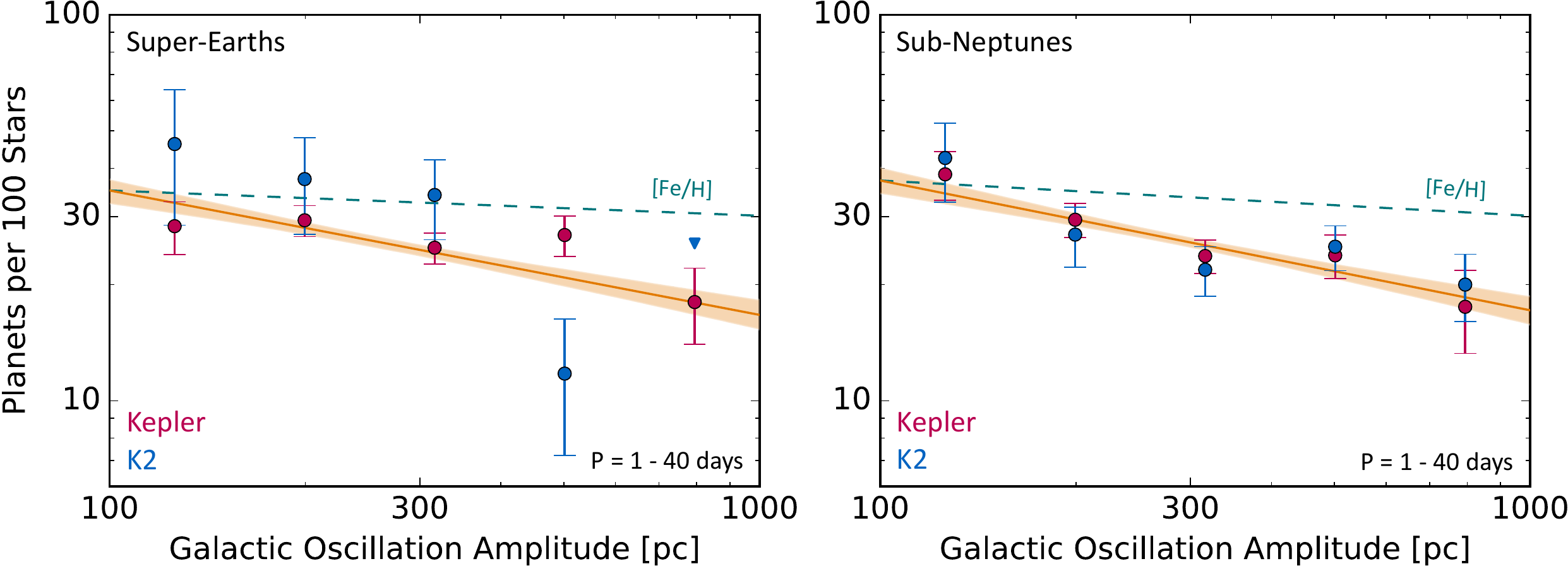}
\caption{ The occurrence of super-Earths and sub-Neptunes as a function of stellar galactic oscillation amplitudes ($Z_\textrm{Max}$). The orange trend lines denote the best-fit model and the $1\sigma$ confidence interval. The teal dotted trend illustrates the occurrence expectation from our calculated galactic [Fe/H] gradient. The triangle shapes represent $3\sigma$ upper limits for each bin. We note that the displayed binned \Kepler and \Ktwo occurrence values are meant to portray the underlying raw occurrence. The model parameterization was performed within our forward-modeling framework using a bin-less CDF optimization. 
\label{fig:galTrend}}
\end{figure*}

\section{Discussion}
\label{sec:Disc}

\subsection{Small Planet Differences}
\label{sec:break}

Our combined analysis of the \Kepler and \Ktwo planet samples used an empirically derived super-Earth and sub-Neptune boundary. Doing so, we found the occurrence of super-Earths and sub-Neptunes turns over at $5.6\pm1.1$ and $9.5\pm2.0$ days, respectively. This is a reduction in the period break reported by \citet{pet18Met}, who found $P_{br}=6.5\pm^{1.6}_{1.2}$ and $P_{br}=11.9\pm^{1.7}_{1.5}$ days for the super-Earth and sub-Neptune populations respectively, when drawing a strict radius partition ($1.7 R_\Earth$). Using a functional class boundary provides a more accurate classification of these small planets and brings the population $P_{br}$ values in closer alignment. The turnover difference in these two populations is expected due to mass loss mechanisms, which remove H/He atmospheres from sub-Neptunes, increasing the short-period super-Earth population. 

It is also notable that the occurrence of super-Earths decreases beyond this break. Previous work found a negligible slope beyond 10 days due to heavy contamination from small-radius sub-Neptunes (i.e., \citealt{pet18Met}). Our upper radius valley and lower $1R_\Earth$ bound may also be responsible for this drop, narrowing the parameter space available for the long-period population. If sample completeness was significant below $1R_\Earth$ and we extended our definition of super-Earths to include sub-Earths, we may expect the long-period occurrence slope to align with the sub-Neptune population. However, \cite{qia21} found an inflection around $1R_\Earth$, suggesting sub-Earths represent a unique population which are not an extension of super-Earths. These planets are likely born intrinsically rocky, and are not the outcome of atmospheric mass loss \citep{owe18,nei20}. Therefore, the occurrence of planets born as sub-Neptunes, which underwent atmosphere erosion, does indeed appear to drop at longer periods where mass loss mechanisms decay in efficiency.

In Figure \ref{fig:metalCompare_long} the warm super-Earth and sub-Neptune populations appear to be independent of stellar metallicity. Furthermore, warm super-Earths present an insignificant correlation with spectral class, while warm sub-Neptune occurrence decreases around earlier-type stars. \citep{pet22} found increasing sub-Neptune radii around more-massive stars, while super-Earths maintain uniformity across a wide stellar mass range. Assuming a constant critical core mass threshold, where planets with cores below this limit undergo photoevaporation, a single core distribution can reproduce this population difference if more-massive stars produce a wider dynamic range of planetary cores (see Figure 15 of \citealt{pet22}). Our stellar independence for warm super-Earths and spectral dependence for warm sub-Neptunes is in alignment with this model. Warm super-Earth production appears to saturate around FGK stars, where sufficient material is available for the primordial core distribution to exceed the critical core mass threshold. In contrast, sub-Neptunes likely undergo luminosity based mass loss and leave behind thick H/He envelopes around only the highest mass portion of the core population. 

Overall, super-Earths present a flattened distribution in log-period space. If these thin envelope planets provide a more pristine portrayal of the underlying core distribution, this population compression provides further evidence of a log-uniform period core distribution, as suggested by unified primordial core models (i.e., \citealt{rog21}). Despite the differences observed between these two populations, it appears that mass loss remains a competent theory for explaining deviations, under the guise of a single birth population.

\subsection{Small Planets and a Common Birth Population}
\label{sec:subsup}
Previous demographic work has shown that a majority of super-Earths hail from a core population common with that of sub-Neptunes \citep{rog21}. If correct, we should expect stellar trend similarities when marginalizing over the full range of periods where mass loss mechanisms are relevant. In other words, if both populations form from a single primordial core distribution, the initial core draw should be consistently dependent on the intrinsic system characteristics. Fortunately, the \Ktwo sample probes exactly the relevant range ---beyond 40 days sub-Neptune mass loss is negligible. In Figure \ref{fig:stellarCompare} we show consistency between these two planet classes and their association with spectral class. However, a trend reduction ($\Delta \gamma=-0.08\pm0.03$) is noted between the super-Earth and sub-Neptune populations, providing tentative evidence for a stellar-driven mass loss mechanism, in contrast with primordial super-Earth formation \citep{lee21}. \citet{rog23} made a similar occurrence ratio argument using stellar mass and found a consistent trend. Alternatively, this difference may highlight a more complex primordial core distribution.

In Figure \ref{fig:supsubCompare} a significant deviation in these two planet classes is observed for periods within 10 days. Here, atmospheric mass loss mechanisms, which are dependent on envelope opacity and metallicity (see \citealt{owe18}), are largely at play. This expected population difference is reflected in the metallicity trend sub 10 days (Figure \ref{fig:metalCompare}). The super-Earth population depicts the impact of metallicity on the primordial core production, while the increased metal trend in the sub-Neptunes highlights the improved atmosphere retention in metal-rich envelopes. The [Fe/H] correlation for these two classes flattens beyond 10 days, but \teff remains an important parameter in warm sub-Neptune occurrence as discussed in Section \ref{sec:break}.

We also find similarity in the galactic amplitude trends (super-Earths: $\tau= -0.30\pm0.06$; sub-Neptunes: $\tau=-0.36\pm0.07$), suggesting the galactic component responsible for this trend is agnostic to sub-Neptunes over super-Earths. The summation of these stellar trends points at a common core birth population.

\subsubsection{Warm Sub-Saturn Origins}
\label{sec:warmSaturns}
The origin of sub-Saturns remains mysterious, as their similar-period integrated occurrence with Jovian planets is in tension with standard core accretion planet formation \citep{pol96}. In Figures \ref{fig:supsubCompare} and \ref{fig:subjupCompare}, the overall shape of the sub-Saturn period population mimics that of the sub-Neptune distribution, with a sharp occurrence increase up to 10 days followed by a slight reduction. This excess of warm sub-Saturns is not replicated in the Jovian population, indicating some unique formation pathways. Furthermore, stellar spectral class appears to play a role in sub-Saturn planet occurrence, whereas no apparent dependence is observed for the Jovian population (see Figure \ref{fig:stellarCompare}). Perhaps this population of warm sub-Saturns is the tail end of the sub-Neptune core distribution. Recent RV follow-up studies have found these warm sub-Saturns harbor a range of eccentricities (e.g., \citealt{now20}); thus, it seems plausible that the most massive sub-Neptune cores may have undergone some tidal radius inflation \citep{mill20}. 

We found consistency of roughly $1\sigma$ in our combined period population models ($\beta_1$,$P_\textrm{br}$, and $\beta_2$) for the sub-Neptune and sub-Saturn planet classes, providing evidence for an interconnected history. Furthermore, we observed a heightened power law dependence for the sub-Saturn radius population ($\alpha=-2.7\pm0.6$) when compared to the sub-Neptunes ($\alpha=-1.7\pm0.1$). If these two planet classes share a common origin, tidal inflation would skew the underlying radius distribution towards smaller planets. In other words, if we assume core masses over some threshold ($\sim10M_\Earth$) undergo runaway accretion and form Jovian planets, cores short of this threshold may experience tidal inflation. This radius enhancement creates a steep population decline in the radius distribution beyond the corresponding mass limit (see Figure 5 of \citealt{mill20}).

Simultaneously, the properties of some sub-Saturns cannot be rectified with tidal inflation; for example, Kepler-1656b is a $5R_\Earth$ planet with a mass of $48\pm4M_\Earth$ \citep{bra18}. These more-massive sub-Saturns exhibit heightened eccentricity and are usually found in single-planet systems \citep{pet17a}, suggesting a planet-planet scattering or merger-based origin after disk dispersal. If the most massive sub-Saturns form through collisions of smaller planet cores, the successor distribution may loosely replicate the underlying progenitor population (i.e., sub-Neptunes) with a normalization reduction dictated by the merger efficiency. The consistency observed between the sub-Saturn and sub-Neptune planet population models, and the reduced overall occurrence of sub-Saturns, remains consistent with this formation mechanism. 

Our population models suggest less-massive sub-Saturns are an extension of the sub-Neptune population. If the most massive sub-Saturns, which have a unique eccentricity and multiplicity distribution, are born through collisions of smaller planets after disk dissipation, our population model also remains consistent with this mechanism. Thus, these two independent formation processes are congruent and yield a coherent population trend that mimics the sub-Neptune population, as observed in our model. Alternatively, giant impacts during planet formation may be sufficient in removing large gaseous envelopes, hampering runaway accretion \citep{bie19}. Nevertheless, it is apparent that the sub-Neptune and sub-Saturn planet populations have an interconnected formation history.

\subsection{Possible Causes of the High Galactic Oscillation Amplitude Deficit}
\label{sec:gaxForm}

In Section \ref{sec:GalAmp} we found a decrease in super-Earth and sub-Neptune occurrence around stars with large galactic oscillation amplitudes. This result is in alignment with \cite{chen21}, who found a reduction in planet multiplicity in thick disk stars. Our trend captures a smooth transition between the thin and thick disk populations, circa 300 pc. Either these two stellar populations are not distinct \citep{par21}, or they have undergone thorough mixing \citep{buc20} as put forth by galactic simulations.

Regardless of the galactic sub-structure origin, high amplitude stars have undergone some additional dynamic heating. Kinematic studies of sub-structure populations observed a 25\% binarity increase in the thick disk compared to thin disk stars \citep{niu21}, pointing to enhanced fragmentation in older metal poor star forming clouds \citep{tan14}. If these high amplitude stars were born in dense stellar clusters, increased dynamical interactions may play a role in the reduced planet occurrence. However, the period range relevant to this sample is deep within the stellar gravitational potential, requiring a very close flyby encounter or some complex instability triggered by the perturbation of an outer giant planet.

The disk element inventory of thick disk stars contains a definitively increased relative alpha-element abundance as compared with thin disk stars. Mg and Si are abundant alpha-elements in terrestrial solar system planets and have condensation temperatures comparable to Fe \citep{lod03}, making them important components of dust in planet-forming regions (e.g., \citealt{gon09}). Additionally, \citet{adi12} found a correlation between planet occurrence and [Ti/Fe] abundances, suggesting Fe-poor stars could still efficiently produce planets so long as the Ti abundance was sufficiently high. Within the \Kepler sample, \citet{bre18} showed that the relative occurrence of small multiplanet systems increases in the low [Fe/H] stellar population, indicating non-Fe elements are largely responsible for their formation. Despite these expected correlations with planet occurrence and alpha abundances, we found that the total population of small planets is reduced in the thick disk where alpha-elements are more dominant. Counter-intuitively, an excessive formation rate could lead to large-scale instabilities that eject most of the small short-period planets \citep{gol22} and leads to an overall reduction in occurrence for the alpha-element rich systems. It may also be the case that the total metal inventory in these high-amplitude stars is lower than thin disk stars, making formation inefficient. In other words, thick disk stars on average only have 50\% ($[\alpha/H]\sim-0.3$ dex) the total alpha-element abundance of thin disk star, despite their dominance over Fe ($[\alpha/Fe]\sim0.3$ dex). If these elements are key to small planet emergence, this deficit may throttle their formation in the thick disk. 

Our occurrence measurements assume an isotropic distribution of inclinations. Perhaps the high-amplitude stars have some preferential alignment with the galactic plane. This would manifest in an occurrence deficit (or surplus depending on the direction of the preference) in our calculated values despite a consistent underlying population. While it is difficult to completely rule out this scenario, the nearby binary population suggests no such inclination anisotropy \citep{aga15}. Therefore, it is unlikely that high-amplitude planetary systems harbor an orbital preference.

Looking at the giant planet population may provide hints as to the root cause of this deficit. Unfortunately, our limited sample of astrometrically resolved giant hosts (38 sub-Saturns and 42 Jupiters) does not yield a meaningful trend in the $Z_{\textrm{Max}}$ axis. It may be the case that these gas-giants are not impacted by the underlying mechanism responsible for the reduced occurrence of super-Earths and sub-Neptunes. Alternatively, these planets may play a role in the absence of smaller planets (i.e., through dynamic instabilities). Without a larger statistical sample of giant planets, it is difficult to determine their significance in this process. Large missions like \tess \citep{tess} and \emph{PLATO} \citep{plato} will provide additional homogeneously identified planet samples that can resolve any existing trends in these giant populations.

\section{Summary and Conclusions}
\label{sec:Conc}
We provide a summary of the work presented in this study:
\begin{itemize}
    \item Here, we carried out a homogeneous analysis of the \Kepler and \Ktwo planet population around FGK dwarfs. In concert, we provided spectroscopic updates to 310 \Ktwo targets using Keck/HIRES, refining the parameters of the underlying stellar population. Overall, we found consistency across all four planet classes (super-Earth, sub-Neptunes, sub-Saturns, and Jupiters). The \Ktwo fields span a much wider portion of the sky, testing various regions of the local galaxy. Analogous occurrence across these fields and the \Kepler postage stamp indicates relatively homogeneous planet occurrence across the local galaxy, further proving the robustness of the \Kepler results. 
    
    \item This work looked at a range of planet classes and aimed to understand the underlying formation mechanisms that carve out each planet population. Standing on the shoulders of previous demographics work, we separated super-Earths and sub-Neptunes along their expected mass loss transition. This careful planet classification led to a more flattened period distribution for the super-Earths, aligning with expectations from primordial core models. 
    
    \item Testing key results from \Kepler, we strengthen trends in stellar spectral class and metallicity. We found super-Earths, sub-Neptunes, and sub-Saturns all diminish in occurrence at higher stellar $T_\mathrm{eff}$,  consistent with previous results, suggesting an inverse formation scaling with disk mass. A flat spectral dependence is found for the Jupiter planet class, indicating some disk mass formation saturation in the FGK mass regime. We also observed consistency with existing metallicity trends, with minor changes bringing the small and giant planet correlations respectively in alignment. The consistency in stellar trends for the sub-Neptunes and super-Earths provide further evidence that these two planet populations were born out of the same core-mass distribution, which underwent apparent envelope mass loss.

    \item We observed a 3$\times$ increase in the occurrence of sub-Saturns relative to Jupiters beyond 10 days. This suggests a distinct formation history. Since mass loss mechanisms are more potent at short periods, it seems likely that some primordial formation process is responsible for this excess. Our finding also provides further support for a warm Jupiter valley (a deficit or flattening of occurrence between 10--100 day periods). We observed that the sub-Neptune and sub-Saturn period population models are consistent to within $1\sigma$ and that the \teff dependence is consistent to $0.5\sigma$, suggesting some interconnection between the formation of these two populations. 
    
    \item Using \GAIA DR3 proper motion and RV measurements, we find a striking trend in planet occurrence as a function of stellar galactic oscillation amplitude. Moving up in amplitude, the number of super-Earths and sub-Neptunes with periods 1--40 days decreases, suggesting a unique history for the high-oscillation-amplitude planet population. If the galactic amplitude is a proxy for stellar age, as expected by galactic sub-structure modeling, it may be that long-term dynamical instabilities are responsible for the lack of small short-period planets. Additionally, thick disk stars have unique element abundance profiles that may contribute to the detected trend. Regardless of the specific dynamics or formation processes responsible for this trend, it is clear that galactic features are imprinted on the planet population. Mapping these galactic influences will provide more robust demographics and a better understanding of our place in the galaxy.  The forthcoming Roman space telescope will attain high cadence photometry for different parts of the local galaxy, with a focus on the galactic bulge. This mission will search a poorly constrained population of planets, which orbit stars near the center of our galaxy. Here, the stellar metallicity gradients (both radially and vertically) are steeper and the stellar density is heightened, potentially modifying the natal disk composition and each system's dynamic history. The results of this survey will provide a more refined understanding of the interplay between planets and the galaxy they inhabit. 

\end{itemize}

\acknowledgments
JZ acknowledges support provided by NASA through Hubble Fellowship grant HST-HF2-51497.001 awarded by the Space Telescope Science Institute, which is operated by the Association of Universities for Research in Astronomy, In., for NASA, under the contract NAS 5-26555. 

This material is based upon work supported by the National Aeronautics and Space Administration under Agreement No. 80NSSC21K0593 for the program “Alien Earths”. The results reported herein benefited from collaborations and/or information exchange within NASA’s Nexus for Exoplanet System Science (NExSS) research coordination network sponsored by NASA’s Science Mission Directorate.

This research has made use of the NASA Exoplanet Archive, which is operated by the California Institute of Technology, under contract with the National Aeronautics and Space Administration under the Exoplanet Exploration Program.
\appendix

\begin{deluxetable*}{lccccccc}
\tablecaption{A list of HIRES/\texttt{isoclassify} stellar parameters updates \label{tab:stellarParams}}
\tablehead{\colhead{EPIC} & \colhead{$T_{\textrm{eff}}$} & \colhead{$\log g$} & \colhead{[Fe/H]} & \colhead{$M_\star$} & \colhead{$R_\star$} & \colhead{$\rho_\star$} & \colhead{Host}\\
\colhead{} & \colhead{K.} & \colhead{dex.} & \colhead{dex.} & \colhead{$M_\Sun$} & \colhead{$R_\Sun$} & \colhead{$g cc^{-1}$} & \colhead{}}
\startdata
201295312 & 5837 & 4.06 & 0.18 & 1.17 & 1.55 & 0.31 & 1	\\					
201338508 & 4066 & 4.70 & -0.47 & 0.56 & 0.55 & 3.26 & 1 \\					
201345483 & 4352 & 4.58 & 0.24 & 0.73 & 0.72 & 1.93 & 0 \\					
201357835 & 5783 & 4.30 & -0.43 & 0.91 & 1.13 & 0.62 & 0 \\					
201384232 & 5692 & 4.55 & -0.10 & 0.92 & 0.89 & 1.27 & 1 \\
\enddata
\tablecomments{This table is available in its entirety in machine-readable form.}
\end{deluxetable*}

\section{New HIRES spectroscopy for \texorpdfstring{$K2$}{K2} targets}
\label{sec:hires}
Our new sample of HIRES spectra are derived following the procedure of the California-\Kepler Survey. We provide a brief summary of the processing undergone to attain our sample of stellar parameters, but suggest interested readers reference \citet{pet17Cat} for a detailed account.

All 456 \Ktwo targets were observed using HIRES on Keck \citep{vogt94} over the course of 2014--2023. This nearly decade-long survey maps well to the release of \Ktwo campaigns as part of the California Planet Search (CPS; \citealt{how10}), with recent additions following the release of the Scaling \Ktwo homogeneous planet catalog \citep{zin21}. The goal of this release is to provide a homogeneous sample of \Ktwo spectra analogous to that of the CKS sample. Our quality requirements are $\textrm{SNR}\ge 45 \,\textrm{px}^{-1}$ with a corresponding $R\ge60,000$, achieved using the "C2" slit. 

Processing of these spectra was done using SpecMatch-synthetic \citep{pet15} for stars with $T_\textrm{eff}>4700$ K (284 stars). This software interpolates over a collection of synthetic spectra in grids of $T_\mathrm{eff}$, $\log g$, and [Fe/H], modifying the model spectra to reflect the instrument profile, stellar turbulence, and stellar rotation. This grid of modified model spectra was then compared against the target spectra using an $L_2$ regularization process, yielding characteristic $T_\mathrm{eff}$, $\log g$, [Fe/H], $v\sin i$ and $M_{\star}$ values. For stars with $T_\textrm{eff}<4700$ K (26 stars), the synthetic models fail to capture the complex molecular features that arise in real spectra, reducing the software's accuracy. For these stars, we instead used SpecMatch-empirical \citep{yee17}, which uses a set of 404 real spectral standards with well-defined \teff, $R_{\star}$, and [Fe/H] parameters to create the model grid. This empirical library captures the complexities that arise in nature, avoiding the model mismatch identified for these cooler stars. The same $L_2$ regularization process was carried out as SpecMatch-synthetic, producing empirically derived \teff and [Fe/H] parameters. 

For stellar mass, radius, age, and density parameters, we rely on isochrone grid matching using spectral characterization, proper motion measurements, photometry, and galactic 3D dust maps. For this optimization, we implemented the \texttt{isoclassify} \citep{hub17,ber20} software, which uses the MESA MIST stellar track models \citep{choi16} to determine the best-fit stellar age, mass, radius, and density. Within this code, the \citet{gre19} dust map was used to account for interstellar extinction, a necessary parameter for luminosity and stellar radius determination. This processing is in line with previous CKS catalogs and encapsulated 73 targets from \citet{pet18c}, which we update with \GAIA DR3 astrometry to provide the most accurate radius parameterization. Overall, these improved proper motion measurements yield stellar radius uncertainties of $\sim2\%$, in agreement with other \GAIA DR3 based parameterization \citep{ber23}. This catalog of stellar parameters includes 239 planet-hosting targets included in the homogeneous Scaling K2 planet candidate table. We provide a list of these updated parameters in Table \ref{tab:stellarParams}.

\section{Normalization of Nonhomogeneous Samples}
\label{sec:norm}
In order to preserve the purity of our stellar parameters, we selected stars with precise spectroscopic and astrometric parameters in Section \ref{sec:trends}. This sample selection is skewed toward bright stars for which high-resolution spectra and proper motion measurements are feasible to obtain. However, these cuts may not be done uniformly. For example, planet-hosting stars, known \textit{a priori}, may be observed with greater frequency than a random sample would otherwise dictate. In previous work, such as \cite{pet18Met}, these selection effects were less important since their largest contributing issue is in the total sample normalization. However, these selection effects impact the \Kepler and \Ktwo samples differently, leading to potential systematic occurrence offsets.

To first order, these sample cuts will impact the occurrence normalization. In other words, the sample reduction may reduce the number of host and non-host systems in a non-parallel fashion, changing the extracted planet occurrence rates. To correct for this offset, we re-normalized the overall occurrence to conserve the full population rates:

\begin{equation}
\kappa=\frac{N_{\textrm{Planets}}}{\eta \; N_{\textrm{Stars}}}  \cdot \frac{\eta^* \; N^*_{\textrm{Stars}}}{N^*_{\textrm{Planets}}}.
\end{equation}
Here, $N$ represents the total number of either planets within a class or stars in the sample, and $\eta$ is the sample completeness. The $^*$  superscript indicates the original sample as discussed in Section \ref{sec:stellarSample} and \ref{sec:planetSample}, while the lack of $^*$ indicates the reduced samples discussed in Section \ref{sec:trends}. Multiplying the appropriate occurrence correction, $\kappa$, by Equation \ref{eq:trend} re-normalizes the reduced catalog to the full sample occurrence. The respective $\kappa$ values used are provided in Table \ref{tab:norm}.

This first-order correction may not be sufficient if the sample reduction artificially skews the stellar populations. Examination of the \teff and stellar radius distributions in Figures \ref{fig:spec_stellar} and \ref{fig:astro_stellar} shows no significant differences with respect to the parent population (Figure \ref{fig:stellarcomp}), indicating the underlying distribution remains intact. Further evidence is in the strong alignment with the \Kepler and \Ktwo samples, which would be unlikely if significant parameter biases existed in either of the population samples. 

\begin{deluxetable*}{lcccc}
\tablecaption{ Table of occurrence normalization corrections ($\kappa$).   \label{tab:norm}}
\tablehead{\colhead{Class} &\colhead{Trend} & \colhead{Period (d)} & \colhead{Mission} &\colhead{$\kappa$}}
\startdata
SE& & & & \\
&[Fe/H] & 1--10 & \Kepler & 0.59\\
&[Fe/H] & 1--10 & \Ktwo & 0.91\\
&[Fe/H] & 10--40 & \Kepler & 0.67\\
&[Fe/H] & 10--40 & \Ktwo & 0.61\\
&$Z_\textrm{Max}$ & 1--40 & \Kepler & 1.0\\
&$Z_\textrm{Max}$ & 1--40 & \Ktwo & 1.5\\
SN& & & &  \\
&[Fe/H] & 1--10 & \Kepler & 0.48\\
&[Fe/H] & 1--10 & \Ktwo & 0.45\\
&[Fe/H] & 10--40 & \Kepler & 0.50\\
&[Fe/H] & 10--40 & \Ktwo & 0.32\\
&$Z_\textrm{Max}$ & 1--40 & \Kepler & 1.0\\
&$Z_\textrm{Max}$ & 1--40 & \Ktwo & 0.68\\
SS& & & & \\
&[Fe/H] & 1--10 & \Kepler & 0.71\\
&[Fe/H] & 1--10 & \Ktwo & 0.50 \\
&[Fe/H] & 10--40 & \Kepler & 0.45\\
&[Fe/H] & 10--40 & \Ktwo & 1.1\\
J& & & & \\
&[Fe/H] & 1--10 & \Kepler & 0.48\\
&[Fe/H] & 1--10 & \Ktwo & 0.66 \\
&[Fe/H] & 10--40 & \Kepler & 0.63\\
&[Fe/H] & 10--40 & \Ktwo & 1.0\\
\enddata
\end{deluxetable*}

\bibliography{paper}{}
\bibliographystyle{aasjournal}

\end{document}